\documentclass[11pt, letterpaper]{article}

\usepackage{graphicx}
\usepackage{amsmath}
\usepackage{amssymb}
\usepackage{setspace}
\usepackage{epstopdf}
\usepackage{color}
\usepackage[letterpaper,left=1in,right=1in,top=1in,bottom=1in]{geometry}
\usepackage[]{algorithm2e}
\usepackage{multirow}
 \usepackage{longtable}
 \usepackage{rotating}
 \usepackage{empheq,etoolbox}

\usepackage{titlesec}

\titleformat{\paragraph}
{\normalfont\normalsize\bfseries}{\theparagraph}{1em}{}
\titlespacing*{\paragraph}
{0pt}{3.25ex plus 1ex minus .2ex}{1.5ex plus .2ex}

\allowdisplaybreaks

\title{\Large \bf A comparative study of model approximation methods applied to economic MPC}
\author{
\centerline{\normalsize Zhiyinan Huang$^{a,*}$, Qinyao Liu$^{a,b,}$\footnote{Both Zhiyinan Huang and Qinyao Liu are co-first authors and have contributed equally to the work. Qinyao Liu focused more on traditional model approximation and identification while Zhiyinan Huang focused more on explicitly-expressed neural network and optimization.}

,\ Jinfeng Liu$^{a,}$\thanks{Corresponding author: J. Liu. Tel:
+1-780-492-1317. Fax: +1-780-492-2881. Email: jinfeng@ualberta.ca.},\ Biao Huang$^{a}$}
 
\vspace{5mm}\\
\centerline{\small $^{a}$Department of Chemical \& Materials Engineering, University of Alberta,}\\
\centerline{\small Edmonton, AB T6G 1H9, Canada}\\
\centerline{\small $^{b}$School of Internet of Things Engineering, Jiangnan University,}\\
\centerline{\small Wuxi  214122, PR China}

}
\allowdisplaybreaks

\begin{document}

\date{}
\maketitle
%\doublespacing
%\onehalfspacing
\setstretch{1.39}

\begin{abstract}
	
	Economic model predictive control (EMPC) has attracted significant attention in recent years and is recognized as a promising advanced process control method for the next generation smart manufacturing. It can lead to improving economic performance but at the same time increases the computational complexity significantly. Model approximation has been a standard approach for reducing computational complexity in process control. In this work, we perform a study on three types of representative model approximation methods applied to EMPC, including model reduction based on available first-principle models (e.g., proper orthogonal decomposition), system identification based on input-output data (e.g., subspace identification) that results in an explicitly expressed mathematical model, and neural networks based on input-output data. A representative algorithm from each model approximation method is considered. Two processes that are very different in dynamic nature and complexity were selected as benchmark processes for computational complexity and economic performance comparison, namely an alkylation process and a wastewater treatment plant (WWTP). The strengths and drawbacks of each method are summarized according to the simulation results, with future research direction regarding control oriented model approximation proposed at the end.	
	
\end{abstract}

\section{Introduction}

% calculation in control and model approximation method
In the past decade, economic model predictive control (EMPC) is recognized as a promising advanced process control and decision-making framework for the next generation smart manufacturing \cite{ELCBook}. However, solving the EMPC optimization problem often leads to significant computational load, especially in large-scale systems \cite{ELCBook,Zeng2015_IECR}. Model approximation has been considered to be an effective way to reduce the computational burden and improve the applicability of EMPC for large-scale processes \cite{Katz2018_CERD}. It aims to approximate the original system with an appropriately simplified model that captures the main dynamics of the original system. The popular methods for model approximation can be broadly classified into three categories: model reduction based on first-principle models (e.g., proper orthogonal decomposition) \cite{Nguyen2020_CCE}, traditional data-based system identification (e.g., subspace identification) \cite{Mckelvey1996_IEEE_TAC,Felici2007_auto,Zhang2019_P,Zhang2019_JPC}, and machine-learning-based model identification \cite{Saki2020_ISAT}.

%\textcolor{red}{Make sure the references are cited in order 1, 2, 3 ... }
% and data-driven methods. Since the first principle modeling requires physical laws of the underlying system, most works are done on model reduction methods and data-driven methods. In the control design of energy-efficient buildings, Robillart et al. used the balanced truncation method to reduce the model order with control strategies designed based on the reduced model \cite{Robillart2019_JPC}.

% based on the model
Model reduction techniques require the existence of a (typically first-principle) model of the considered system \cite{Acle2019_IEEE_TPS,Ibrir2018_Auto}. Many model reduction methods in this category are projection-based methods, in which the original high-order system is projected onto a lower-order reduced space \cite{Ibrir2018_Auto}. Among them, the proper orthogonal decomposition (POD) is one of the most widely used methods, which is for example applied in \cite{Nguyen2020_CCE} together with the discrete empirical interpolation method to solve large-scale model predictive control (MPC) problems for continuous chemical processing.

% system identification

In the cases that only the input-output dataset is available, system identification techniques are frequently used to establish the mathematical model of a dynamical system \cite{Diao2018_Auto, Liu2020_AMM}, where relatively simple models can be generated without knowledge regarding the underlying physical laws of the system. There are mainly three types of system identification methods, namely the prediction error method  \cite{Anderson1978_Auto}, maximum likelihood method \cite{Akaike1973_B}, and subspace identification method \cite{Mckelvey1996_IEEE_TAC}. Different from the first two methods that require pre-selected model structure and solving optimization problems, the subspace identification method can directly extract the state-space model from the input-output data without using optimization techniques \cite{Mckelvey1996_IEEE_TAC, Zhang2019_JPC}. 
%By employing the constrained least squares to incorporate the prior information, a close-loop subspace identification approach was proposed to improved the accuracy and stability \cite{Zhang2019_JPC}.

% neural network

Different from system identification methods, data-driven machine-learning-based model identification typically leads to neural network (NN) models that do not have explicit mathematical expressions \cite{Saki2020_ISAT}. With the ability to acquire knowledge from the input information and remember the learning results, neural networks can approximate any continuous function according to the universal approximation theorem \cite{Vargas2019_Neuro}. Much attention has been attracted to neural-network-based system modeling \cite{Xu2020_Auto, Koronaki2020_CEJ} and control \cite{Wu2020_JPC}. However, for various applications, neural networks often suffer from their black-box nature. Thus explicitly expressed NNs were proposed and applied in some areas \cite{Zhou2020_CS}.

% Our work

In this work, we perform critical examination of these model approximation methods when used in conventional tracking MPC and EMPC. One representative model approximation method is selected from each of the three model approximation categories. Specifically, POD integrated with trajectory piecewise linearization (TPWL) \cite{Zhang2019_P}, subspace identification \cite{Zhang2019_JPC}, and explicitly-expressed fully connected neural network is selected for this work. Two processes that are very different in dynamic nature and complexity are selected as benchmark processes. Conventional tracking MPC controllers and EMPC controllers designed based on the three model approximation methods are applied to the benchmark processes. The performance of the control framework is compared from different aspects, including computational complexity, tracking performance for MPC, and economic performance for EMPC.
%To the best of our knowledge, this is the first work that applies neural networks as system models in MPC and EMPC controller design for high dimensional nonlinear systems. 
 % are selected the three categories (one from each) and are used in the design of EMPC controllers for a large-scale wastewater treatment plant (WWTP).  The core ideas and brief procedures of each method are discussed, followed by EMPC simulation results obtained with a wastewater treatment plant (WWTP). The EMPC performance achieved with the original models is employed as the reference point of performance comparison. Remarks are provided based on the case study result, with the benefits and drawbacks of each method summarized.

% paper structure

% Briefly, the paper is structured as follows. Section~\ref{Preliminaries} presents some preliminaries including the EMPC design employed. In Section~\ref{MAM}, the selected model approximation algorithms are introduced. Then in Section \ref{Results} the WWTP process is introduced, followed by the simulation results and some remarks regarding the proposed EMPC frameworks. Finally, conclusions and plans for future works are given in Section~\ref{Conclusions}.

\section{Preliminaries}
\label{Preliminaries}

\subsection{System description and problem formulation}

In this work, we consider model approximation of nonlinear dynamical systems that can be described using the following ordinary differential equations:
\begin{subequations}\label{eqn:sys}
	\label{sys_o}
	\begin{empheq}{align}
		\label{lqy2_1}
		\dot{x}(t)&=f(x(t),u(t))\\
		\label{lqy2_2}
		y(t)&=g(x(t),u(t))
	\end{empheq}
\end{subequations} 
where $x\in\mathbb{R}^{n}$ is the state vector, $u\in\mathbb{R}^{r}$ is the input vector, $y\in\mathbb{R}^{l}$ is the output vector, $f$ and $g$ are nonlinear vector functions, $f$ describes the dynamics of the system and $g$ describes the relation of the output to the state and the input. For convenience, we will refer to the system in (\ref{eqn:sys}) as the {\em original system} in the remainder of this work. 

Model approximation will be applied to either reduce the original system to a system of a lower order that approximates the dynamics of the original system with sufficient accuracy or to identify a linear or neural network model based on the input-output data of the original system. These reduced or identified models will be applied in advanced control frameworks as system constraints and their applicability and performance will be carefully examined. To be specific, conventional set-point tracking MPC and EMPC will be investigated and differences in control behaviors of these two frameworks with respect to different reduced or identified models are of our interest.

\subsection{Model predictive control design}

We consider a MPC formulation that tracks an optimal steady-state operating point $(y_s,u_s)$. The tracking MPC at a specific sampling time $t_i$ is formulated as an optimization problem as follows:

\begin{subequations}
	\label{MPC}
	\begin{empheq}{align}
		\label{MPC1}
		\min_{u(t)\in S(\triangle)} \; &\sum_{j = i}^{i+N-1} \Big(|y(t_j|t_i) - y_s|^2_Q + |u(t_j|t_i) - u_s|^2_R\Big) +|y(t_{k+N}|t_i) - y_s|^2_{P_f}\\
		\label{MPC2}
		\textmd{s.t. :} \; &\dot{x}(t)=f(x(t),u(t)) \\
		\label{MPC3}
	    &y(t)=g(x(t),u(t))\\
		\label{MPC4}
		&x(t_i) = \bar{x}(t_i) \\
		\label{MPC5}
		&u(t) \in \mathbb{U}\\
		\label{MPC6}
		&y(t) \in \mathbb{Y}
	\end{empheq}
\end{subequations} 
where $S(\triangle)$ is the family of piece-wise constant function with $\triangle$ being the sampling time, $N$ is the prediction horizon and is a finite positive integer, $y(t|t_i)$ and $u(t|t_i)$ represent the prediction of the variables at future time $t$ made at the current time $t_i$. $Q$, $R$ and $P_f$ are diagonal weighting matrices, $\bar{x}(t_i)$ is the estimated or measured state vector at current time $t_i$, $\mathbb{U} \subset \mathbb{R}^r$ and $\mathbb{Y}\subset \mathbb{R}^l$ are compact sets. The optimization problem aims to optimize $u(t)$ from $t_i$ to $t_{i+N-1}$, such that the accumulated difference between $(y(t|t_i),u(t|t_i))$ and the set-point $(y_s,u_s)$ is minimized, which is defined by the objective function (\ref{MPC1}). System model (\ref{MPC2}) - (\ref{MPC3}) are employed to predict $y(t|t_i)$ with initial condition specified by (\ref{MPC4}). (\ref{MPC5})-(\ref{MPC6}) are input and output constraints that have to be satisfied over the entire horizon $N$ while solving the problem. 

Note that when a reduced or identified model is used to design the MPC controller, the model equations (\ref{MPC2})-(\ref{MPC3}) should be replaced with the reduced or identified model equations accordingly. %(\ref{EMPC}) presents a general form of EMPC. is very basic for simplicity, as EMPC design is not the major focus of this work. Additional terms such as terminal cost and constraints can be added accordingly for system closed-loop stability and better control performance.

The optimal solution of the above optimization problem can be denoted as $u^\star_{\text{EMPC}}(t|t_i)$. However, only the values at the first time instant (i.e. the current time $t_i$) will be applied to the system:
\begin{eqnarray}
	u(t) = u_{\text{EMPC}}^\star(t|t_i), \quad t \in [t_i, t_{i+1})
\end{eqnarray}

Regarding the optimal operating point $(y_s, u_s)$ employed in (\ref{MPC1}), it may be calculated through a steady-state optimization problem as shown in (\ref{RTO}),
\begin{subequations}
	\label{RTO}
	\begin{empheq}{align}
		\label{RTO1}
		(y_s, u_s) =  &\min_{y,u} \; l_e(y,u)\\
		\label{RTO2}
		\textmd{s.t.}\ :\; &\dot{x}=f(x,u) \\
		\label{RTO3}
		&y=g(x,u)\\
		\label{RTO4}
		&u \in \mathbb{U}\\
		\label{RTO5}
		&y \in \mathbb{Y}
	\end{empheq}
\end{subequations} 
where the objective function $l_e(y,u)$ represents an economic performance of the system. 
\subsection{Economic model predictive control design}

As have presented in the previous section, MPC takes a two-layered design. The optimal operating point is first obtained by solving a steady-state optimization problem. A series of dynamic optimization problems that tracks the states and inputs to the optimal operating point is then proceeded along the horizon of interest. In EMPC design, these two steps are combined into a single step. The objective function used in the steady-state optimization, which is often times economic performance related, is directly employed in the dynamic optimization problems.

A basic EMPC formulation for the system described in (\ref{sys_o}) is employed in this work. EMPC is in general formulated as an optimization problem, which takes the following form  at the sampling time $t_i$ \cite{ELCBook}:
\begin{subequations}
	\label{EMPC}
	\begin{empheq}{align}
		\label{lqy2_7}
		\min_{u(t)\in S(\triangle)} \; &\sum_{j = i}^{i+N-1} l_e(y(t_j|t_i),u(t_j|t_i)) \\
		\label{lqy2_8}
		\textmd{s.t.}\ :\; &\dot{x}(t)=f(x(t),u(t)) \\
		\label{lqy2_9}
		&y(t)=g(x(t),u(t))\\
		\label{lqy2_10}
		&x(t_i) = \bar{x}(t_i) \\
		\label{lqy2_11}
		&u(t) \in \mathbb{U}\\
		\label{lqy2_12}
		&y(t) \in \mathbb{Y}
	\end{empheq}
\end{subequations} 

The same definitions as used in the MPC formulation are also valid for the EMPC formulation. The only difference lies in the objective function as used in (\ref{lqy2_7}), where the economic performance index employed in the steady-state optimization is adapted in dynamic optimization directly.

\section{Model Approximation Methods}
\label{MAM}
In this section, the model approximation methods encountered in this work are briefly introduced. The characteristics of each approaches are discussed, followed by step-by-step descriptions of how the methods can be implemented. 
\subsection{Model reduction based on first-principle model}
First, we introduce the proper orthogonal decomposition (POD) and trajectory piecewise linearization (TPWL) methods. The two methods represent the two popular approaches to reduce model complexity respectively, namely reducing the model order and mimicking complex mathematical relationships in simpler structures (e.g. linearization). Applications of these two methods can be found in various works, for example, \cite{Zhang2019_P, Yin2018_CCE} for POD and \cite{Zhang2019_P, Tiwary_IEEE} for TPWL. A combination of POD and trajectory piecewise linearization (TPWL) is considered in this work. 

The key idea of the POD method is to obtain a projection matrix that projects the original state vector onto a lower-dimensional space. To achieve this goal, a snapshot matrix of the states $X$ is first generated by simulating the original system with an input trajectory that ensures persistent excitation of the system. The snapshot $X$ contains sampled system states at different time instants. Suppose that:
\[
X=[x(t_1),x(t_2),x(t_3),\cdots,x(t_{N})]\in\mathbb{R}^{n\times N}
\]
where $x(t_i)$, $i = 1, \cdots , N$, is the sampled state vector at time $t_i$, $N$ is the total number of samples, $n$ is the dimension of the state vector. Note that $N$ should be much greater than $n$ to ensure that the snapshot captures the dynamics of the original system. After $X$ is generated, singular value decomposition (SVD) is applied to $X$ to obtain the following decomposition:
\begin{equation}
	\label{SVD}
	X = U \Sigma V^\text{T}
\end{equation}
where $U\in\mathbb{R}^{n\times n}$ and $V\in\mathbb{R}^{N\times N}$ are orthogonal matrices and are referred to as the left and right singular matrix of $X$ respectively, $\Sigma$ = $\text{diag}\footnote{\textmd{diag}(v) denotes a diagonal matrix with the diagonal elements being the elements of the vector v}([\sigma_1$, $\sigma_2$, $\cdots$, $\sigma_n])\in\mathbb{R}^{n\times N}$ is a diagonal matrix, where $\sigma_1, \cdots, \sigma_n$ are the $n$ singular values of $X$ and are arranged in a descending order. It is noted that the greater the value of a singular value, the more important system dynamics it captures. In POD, the order of the reduced system is determined based on these singular values. It is recommended that the order $k$ is obtained such that there is a significant drop in the singular value magnitude from $\sigma_k$ to $\sigma_{k+1}$. Once $k$ is determined, the projection matrix $U_k \in \mathbb{R}^{n \times k}$ is defined to be the first $k$ columns of the matrix $U$. The reduced system state $z(t)$ has the following relationship with the original state $x(t)$:
\begin{equation}
	\label{PODr}
	x(t) = U_k z(t)
\end{equation}

It is to be noted that although POD reduces the system order to a smaller number, the underlying calculations in general do not reduce from a numeric optimization point of view. This is because that it is in general challenging to explicitly express the reduced system for medium to large scale nonlinear systems. Thus when the original system has complex nonlinear dynamics, applying only POD to the system often does not reduce the computational complexity. A common way to deal with this problem is to simplify the mathematical complexity of the original system first, followed by applying POD on the simplified model. Readers can refer to \cite{Zhang2019_P} for an example application of this approach, where POD was applied to a linear time-varying (LPV) reduced model obtained using TPWL. 

Instead of linearizing at a single point, in TPWL, the original system is linearized along a typical state trajectory at multiple points. With a predetermined number of linearization points, the operating region of interest should be covered evenly by the selected linearization points. This can be achieved by the application of a distance-based algorithm which is introduced in \cite{Rewienski2003_IEEE}. After linearization, the linearized models at different points are combined through weighted summation to form an LPV model. The weighting function may be designed in a way such that once the state-input pair approaches one of the linearization points, the weight for the corresponding sub-model increases to one quickly.  
%\textcolor{red}{Explain why POD only does not work and linearization is needed briefly. Also, mention or explain that TPWL is a common linearization method used together with POD. remember to include the supporting references. Improve or correct the following sentences. If linearization is done first, you may want to explain a little bit why}. 

%\textcolor{red}{Correct if you see any mistakes.} 
Let us assume that there are $s$ linearization points. The weighting function $w_i(\cdot), i = 0, \cdots, n-1$ that combines the linearized models can be determined following \cite{Rewienski2003_IEEE}. The procedure for implementing POD with TPWL is summarized as follows\cite{Rewienski2003_IEEE}: 
\begin{enumerate}
	\item Run open-loop simulations using the original model and generate the snapshot matrix $X$ using the open-loop simulation data.
	\item Based on the snapshot matrix $X$, calculate the projection matrix $U_k$ as the truncated left singular matrix $U_k$ of the snapshot matrix $X$ following (\ref{SVD}) and (\ref{PODr}).
	\item Based on extensive simulations of the original system, determine a representative trajectory of the system and along the trajectory, determine the $s$ linearization points based on the distance-dependent algorithm discussed in \cite{Rewienski2003_IEEE}. 
	\item Obtain the $s$ linearized models by linearizing the original system at each of the $s$ linearization points.
	\item Based on the $s$ linearized models and the weighting function $w_i(\cdot), i = 0, \cdots, n-1$, the LPV model that combines all the linearized models takes the following form:
	\begin{subequations}
		\label{TPWL}
		\begin{empheq}{align}
			\label{TPWL_1}
			\dot{x} &= \sum^{s-1}_{i=0} w_i(x)(A_i(x-x_i)+B_i(u-u_i)+f(x_i,u_i))\\
			\label{TPWL_2}
			y &= \sum^{s-1}_{i=0} w_i(x) (C_i(x-x_i)+D_i(u-u_i)+g(x_i,u_i))	
		\end{empheq}
	\end{subequations} 
	where $x_i, u_i$ for $i = 0,1,...s-1$ represents the linearization points, $A_i, B_i, C_i, D_i$ are the partial derivatives $\dfrac{\partial f}{\partial x}$, $\dfrac{\partial f}{\partial u}$ and $\dfrac{\partial g}{\partial x}$, $\dfrac{\partial g}{\partial u}$ evaluated at the linearization points, respectively. 
	
	%, where $k$ is the order of the reduced model and may be determined by observing when the singular values of $X$ drop significantly.
	\item Use the projection matrix $U_k$ to project the high-order LPV model (\ref{TPWL}) onto a lower-order state-space as follows:
	\begin{subequations}
		\label{TPWL_POD}
		\begin{empheq}{align}
			\label{TPWL_POD_1}
			\dot{z} &= \sum^{s-1}_{i=0} w_i(U_k z)U_k^{T}[(A_i(U_k z-U_k z_i)+B_i(u-u_i)+f(U_k z_i,u_i))]\\
			\label{TPWL_POD_2}
			y &= \sum^{s-1}_{i=0} w_i(U_k z) (C_i(U_k z-U_k z_i)+D_i(u-u_i)+g(U_k z_i,u_i))	
		\end{empheq}
	\end{subequations}
	where $z_i$ represents the $i$th reduced linearization point.
	%$A_{ir} = U_k^{\text{T}} A_i U_k$, $B_{ir} = U_k^{\text{T}} B_i$, $\gamma_{ir} = U_k^{\text{T}}(f(x_i, u_i) - A_i x_i - B_i u_i)$, $C_{ir} = C_i U_k$, $\xi_{ir} = g(x_i,u_i) - C_i x_i - D_i u_i$.
\end{enumerate}

%To summarize, the POD method reduces model complexity through projecting the state space onto a lower dimension subspace, while the TPWL method generates linear parameter varying (LPV) models with linear sub-models obtained around multiple pre-selected points. Note that although POD methods could potentially reduce the computational efforts required for solving optimization problems due to the reduced model dimension, the underlying calculations are still carried out using the original model. Thus the POD method on its own is often not ideal for systems with complex nonlinear dynamics. In this work, the POD method is applied to original models along with the TPWL method and the performance is compared with that achieved when only the TPWL method is used. The EMPC optimization problems are solved with a Python-based optimization toolbox designed specifically for MPC, namely the MPCTools toolbox \cite{MPCTools}.

\subsection{Subspace model identification}

When the first-principle model is not available, model identification based on input-output data of the system is the common approach for model development. There are many different model identification algorithms. In this work, we will focus on subspace identification. Unlike other traditional identification methods that require predetermined model structure and order, subspace identification has the unique property that the model can be identified with only the input-output dataset. Furthermore, the subspace method is purely linear-algebra-based and does not involve solving optimization problems \cite{Mckelvey1996_IEEE_TAC}. % Thus instead of having full access to the state variables, only the input-output dataset is available. For the case study, the outputs are selected to be the states shown up in the economic cost function, and the systems are confirmed to be observable with the selected outputs. 

The subspace identification method leads to a discrete-time linear time-invariant (LTI) state-space model in the following form:
\begin{subequations}
	\label{subspace}
	\begin{empheq}{align}
		\label{subspace_1}
		z_{k+1} &= Az_k + Bu_k,\\
		\label{subspace_2}
		y_k &= Cz_{k} + Du_k,
	\end{empheq}
\end{subequations} 
where $z_k\in\mathbb{R}^{s}$, $u_k\in\mathbb{R}^{r}$ and $y_k\in\mathbb{R}^{l}$ are the state vector, input vector and output vector, respectively, $A\in\mathbb{R}^{s\times s}$, $B\in\mathbb{R}^{s\times r}$, $C\in\mathbb{R}^{l\times s}$ and $D\in\mathbb{R}^{l\times r}$ are system matrices. The model is obtained by first calculating the extended observability matrix of the system using techniques such as oblique projection, from which matrices $C$ and $A$ can be calculated. Matrices $B$ and $D$ can then be obtained using least square methods.

The subspace identification algorithm can be summarized as follows\cite{Mckelvey1996_IEEE_TAC}:
\begin{enumerate}
	\item Calculate the oblique projection of $Y_f$ along the row space of $U_f$ on to the row space of $Z_p$, where $Y_f$, $U_f$, and $Z_p$ are the block Hankel matrix consists of future outputs, future inputs and past inputs and outputs ($Z_p = [U_p^T \: \: Y_p^T]^T$, $U_p$ and $Y_p$ are the Hankel matrix consist the past input and outputs) respectively. 
	\begin{equation}
		O_i = Y_f/_{_{U_f}} Z_p
	\end{equation}

	\item Perform SVD on $O_i$:
	\begin{subequations}
		\label{subspace_SVD}
		\begin{empheq}{align}
			\label{lqy4_7}
			O_i &= \left(\begin{array}{cc}
				U_1 & U_2\\\end{array}\right) \left(\begin{array}{cc}
				S_1 & 0 \\
				0   & 0   \\\end{array}\right) \left(\begin{array}{c}
				V^{\text{T}}_1  \\
				V^{\text{T}}_2
				\\\end{array}\right),\\
			\label{lqy4_8}
			\Gamma_i &= U_1 S^{1/2}_1,\\
			\label{lqy4_9}
			X_f &= \Gamma_i^{\dag} O_i,
		\end{empheq}
	\end{subequations} 
	where the columns of $(U_1 \quad U_2)$ are the left singular vectors of $O_i$, $S_1$ is the diagonal matrix that contains the nonzero singular values of $O_i$, $\Gamma_i$ is the extended observability matrix, $X_f$ is the Hankel matrix that consists the future states, $(\cdot)^{\dag}$ denotes the Moore-Penrose pseudo-inverse of the matrix $(\cdot)$. The dimension $s$ of the state is determined by the dimension of $S_1$.
	\item Compute the extended observability matrix $\Gamma_i$ based on
	(\ref{lqy4_8}), and the orthogonal complement of the row space of
	$\Gamma_i$ can be calculated by $\Gamma^{\bot}_i$ =
	$U^{\text{T}}_2$, where $(\cdot)^{\bot}$ denotes the orthogonal complement of the row space of the matrix $(\cdot)$. 
	
	\item Determine $C$ as the first $l$ rows of $\Gamma_i$, where $l$ is the dimension of the output vector. According to the definition of $\Gamma_i$, we can find the following relation:
	\[
	\left(\begin{array}{c}
		C \\
		CA \\
		CA^{2}\\
		\cdots\\
		CA^{i-2} \\\end{array}\right) A = \left(\begin{array}{c}
		CA \\
		CA^{2} \\
		CA^{3}\\
		\cdots\\
		CA^{i-1} \\\end{array}\right), \quad \underline{\Gamma_i}
	A = \overline{\Gamma_i},
	\]
	where $\underline{\Gamma_i}$ denotes $\Gamma_i$ without the last $l$
	rows and $\overline{\Gamma_i}$ is $\Gamma_i$ without the first $l$
	rows. Then $A$ can be solved by $A$=$\underline{\Gamma_i}^{\dag}$
	$\overline{\Gamma_i}$. 
	\item Calculate the matrix $\Gamma^{\bot}_i$$Y_f$$U^{\dag}_f$, and construct
	{{\begin{eqnarray}\nonumber
				\begin{aligned}
					\Gamma^{\bot}_i Y_f U^{\dag}_f &= \left(\begin{array}{cccc}
						M_1 & M_2 & \cdots & M_i
						\\\end{array}\right), \\
					\Gamma^{\bot}_i &= \left(\begin{array}{cccc}
						L_1 & L_2 & \cdots & L_i
						\\\end{array}\right),
				\end{aligned}
	\end{eqnarray}}}
	then $B$ and $D$ can be solved from
	{\[
		\left(\begin{array}{c}
			M_1\\
			M_2\\
			\vdots\\
			M_i\\\end{array}\right) = \left(\begin{array}{ccccc}
			L_1 & L_2 & \cdots & L_{i-1} & l_{i}\\
			L_2 & L_3 & \cdots & L_{i} &  0\\
			L_3 & L_4 & \cdots & 0 & 0\\
			\cdots  &  \cdots   &     \cdots &  \cdots  &   \cdots\\
			L_i &  0  &  \cdots &  0 & 0\\\end{array}\right)
		\left(\begin{array}{cc}
			I_l & 0\\
			0   & \underline{\Gamma_i}\\\end{array}\right)  \left(\begin{array}{c}
			B\\
			D\\\end{array}\right)
		\]}
\end{enumerate}

\subsection{Neural network modeling}
The last method to be introduced is to train neural networks as system models. Similar to the subspace identification method introduced in the previous section, neural network training does not require a significant amount of knowledge regarding the dynamics of the interested system. Furthermore, NNs can capture nonlinear dynamics much easier compared to the subspace identification method. According to the universal approximation theorem \cite{UAR}, neural networks in general are capable of approximating any function $f:\mathbb{R}^n\rightarrow\mathbb{R}^m$ with n-dimension inputs and m-dimension outputs up to some maximum error $\varepsilon$. In this work, fully connected neural networks with nonlinear activation functions are considered. Unlike models obtained with traditional identification methods, neural networks are often used as black-box models without explicit expressions, which could be challenging for advanced control applications. This is because many of the conventional optimization solvers require Jacobian/Hessian information of the systems, which could be computationally expensive for black-box models. In this work, NN system models are investigated in the form of basic black-boxes as well as explicitly expressed systems of equations.

%\begin{figure}
%	\centering
%	\includegraphics[width=1\columnwidth]{RNN.png}
%	\caption{A recurrent neural network diagram with its structure unrolled.}
%	\label{RNN}
%\end{figure}
A brief step-by-step procedure for identifying a NN model is outlined as follows \cite{Wu2019_AIChE}:
\begin{enumerate}
	\item Determine the dimension of the NN input and output vectors:
	\begin{align*}
		NN_{in} &\in \mathbb{R}^{N_ {past} \cdot (r+l) + (N_{future}-1)\cdot r} \\
		NN_{out} &\in \mathbb{R}^{N_{future} \cdot l}
	\end{align*}
	 where $N_{past}$ is the number of past steps considered, $N_{future}$ is the number of future time steps to be predicted. Note that the inputs of the NN includes both system inputs and outputs from the past. Furthermore, when system output at multiple future steps are of interests, the future system inputs for the corresponding steps are as well required as NN inputs.
	\item Determine the number of NN layers, the number of nodes in each layer, and the activation function to be used. Note that there are no universal rules for determining these parameters. Users can manually do guess-and-check or use tools such as grid search.
	\item Reconstruct the dataset according to the predetermined input-output dimension. 
	\item Identify the NN using computational tools (e.g. the state-of-the-art application program interface (API) Keras \cite{Keras}).
	\item Extract the trained parameters from the model.
	% (e.g. use method \texttt{tensorflow.model.trainable\_variable} if Tensorflow \cite{TensorFlow} is used for NN training). 
	\item Define explicit models with the trained parameters and the known NN structure. A single fully-connected NN can be defined as follows:
	\begin{equation}
		y = \sigma(\sum_{i = 1}^{n_{node}}w_i x_i + b_i)
	\end{equation}
	where $x$ and $y$ represents the input and output of the NN layer, $w_i$ and $b_i$ are the weighting factor and the bias for the $i^{th}$ input, $n_{node}$ is the number of nodes in the NN layer, $\sigma(\cdot)$ is the activation function. 

\end{enumerate}

In the following two sections, we consider two benchmark process examples and study the performance of the three model approximation methods when they are used in MPC and EMPC designs. The two benchmark processes are selected to be very different in scales and dynamic natures, such that the performance of the model approximation methods applied to different types of systems can be investigated and a more comprehensive comparison can be done.
%\textcolor{red}{I moved the paragraph you had to the simulations section. Add something else here or above the procedure. Discribe the type of model obtained from RNN - black box and the challenges caused by this. }

%============================ RESULTS ============================
\section{Case Study 1: Benzene Alkylation Process}
\subsection{Process overview}
\begin{figure}[!t]
	\centering
	\includegraphics[width=0.68\hsize]{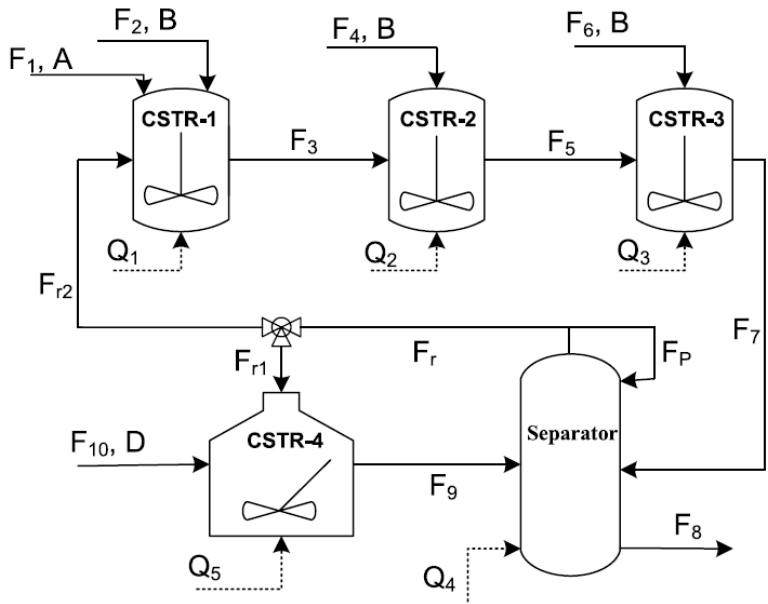}
	\caption{A schematic of the alkylation of benzene process}
	\label{lqy_fig1}
\end{figure}

The first benchmark process considered is a benzene alkylation process. A schematic diagram of the process is presented in Figure~\ref{lqy_fig1} \cite{LiuDMPC}. The process considered consists of four CSTRs and a flash tank separator. Stream $F_1$ consist of pure benzene while streams $F_2$, $F_4$ and $F_6$ consist pure ethylene. Two catalytic reactions take place in CSTR-1, CSTR-2, and CSTR-3, of which benzene ($A$) react with ethylene ($B$) and produce the desired product ethylbenzene ($C$) (reaction 1), while ethylbenzene further react with ethylene and produce the byproduct 1,3-diethylbenzene ($D$) (reaction 2). The effluent of CSTR-3 is fed to the flash tank separator, where the desired product is separated and leave from the bottom stream $F_8$. On the other hand, the left-over benzene leaves the separator from the top and splits into $F_{r1}$ and $F_{r2}$ which goes into CSTR-4 and CSTR-1 respectively. 1,3-diethylbenzene is fed into CSTR-4 through $F_{10}$, where it reacts with benzene to produce ethylbenzene under the presence of catalysts (reaction 3). All the chemicals leaving from CSTR-4 eventually goes to the separator. Each vessel has an external heat supply stream which helps to adjust the vessel temperature ($Q_1$ - $Q_5$).

The state variables of the process are the concentrations of the four chemical components and the temperature in each vessel. The manipulated inputs are the external heat supply and the feed stream flow rates of ethylene going into CSTR-2 and CSTR-3. Overall, the system consists of 25 states and 7 manipulated inputs. A more detailed model description with parameter values can be found in \cite{LiuDMPC}. 

\subsection{Control objective and controller setting}
\label{control setting}

\begin{figure}
	\centering
	\includegraphics[width=0.8\columnwidth]{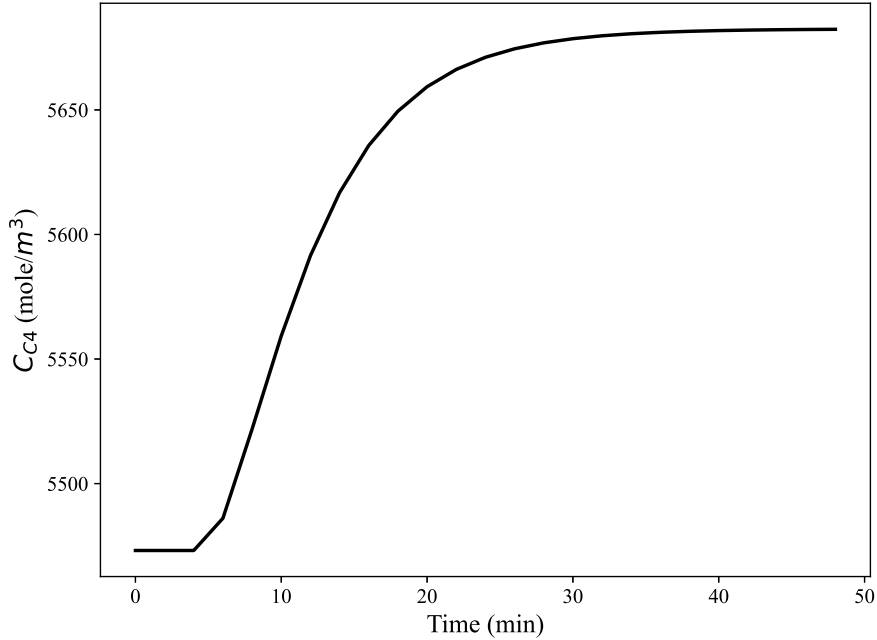}
	\caption{Step response of the alkylation process output.}
	\label{stepResponse_A}
	% \Model Reduction\Alkylation\runMPC
\end{figure}

The control objective is to maximize the production rate of ethylbenzene while keeping the energy cost at a reasonable level. Thus the economic performance measure employed for the steady-state optimization and the EMPC controller is defined as follows:
\begin{equation}
	\label{ecost}
	l_e =  -\alpha F_8 C_{C4} + \beta \sum_{i = 1}^{5} |Q_{is} + Q_i| 
\end{equation}
where $F_8$ and $C_{C4}$ are the flow rate and ethylbenzene concentration leaving the separator respectively, $Q_i$ represents the deviation heat inlet to vessel $i$ with respect to the steady-state values. $Q_{is}$ is the steady-state heat inlet for the corresponding vessel, $\alpha = 200$ and $\beta = 0.0005$ are the weighting coefficients selected so that a fair trade-off between production rate and operating cost is established. Note that although the flow rates of ethylene feeding into CSTR-2 and CSTR-3 do not present explicitly in the objection function, the value of $F_8$ however depends on them.

The weighting factors used in the MPC tracking objective function (\ref{MPC1}) were selected to be $Q = 1$, $R = \textmd{diag}([7.5, 5, 5, 6, 5, 4.9\cdot10^{-4}, 4.9\cdot10^{-4}])$ and $P_f = 10$, where a higher penalty was added to the terminal output to enhance convergence. For simplicity, the output vector $y$ is $C_{C4}$, which is the only state appeared in (\ref{ecost}). The observability of the system is confirmed using the PBH test along the data trajectory with the selected output. The optimal steady-state operating point $y_s = 5526$, $u_s = [1.0,1.0,1.0,-1.0,-0.86,1.0,1.0]$ is obtained by solving the steady-state optimization (\ref{RTO}). The input vector and the output are bounded based on physical properties of the process. 
%The first five inputs are the heat inlet stream going into the 
\begin{subequations}
	\begin{empheq}{align}
		\mathbb{U} &= \{u|\: \underbar{u} \leq u \leq \bar{u}\} \\
		\mathbb{Y} &= \{y|\: y\geq 0\}
		%, \: u_{min} = [-1.0, -1.0, -1.0, -1.0, -1.0, 0.0, 0.0],  \:[1.0, 1.0, 1.0, 1.0, 1.0, 1.0, 1.0]
	\end{empheq}
\end{subequations}
where $\underbar{u}$ and $\bar{u}$ are the upper and lower bound of the input vector. The exact values of $\underbar{u}$ and $\bar{u}$ can be found in \cite{LiuDMPC}.

The controller sampling time is 2 minutes, and the prediction horizon is $N = 15$. The two parameters are picked based on the output response (Figure \ref{stepResponse_A}) with a unit step change in the input vector $u$. As presented in Figure \ref{stepResponse_A}, the major dynamics of the system can be captured in the 30-minute window defined by the sampling time and prediction horizon.

\subsection{Data generation and model approximations}
\label{Data_A}
All the methods discussed in section 3 require the generation of a dataset that can represent the dynamics of the original process. Given that the considered alkylation process is nonlinear, multi-level input signals are used as the inputs to the process for input-output data generation. Open-loop simulations are carried out with the multi-level inputs applied to the original nominal system model. For consistency, one dataset is generated and is used for all the model approximation methods.

\begin{figure}[!t]
	\centering
	\epstopdfsetup{outdir=./}
	\includegraphics[width=0.9\columnwidth]{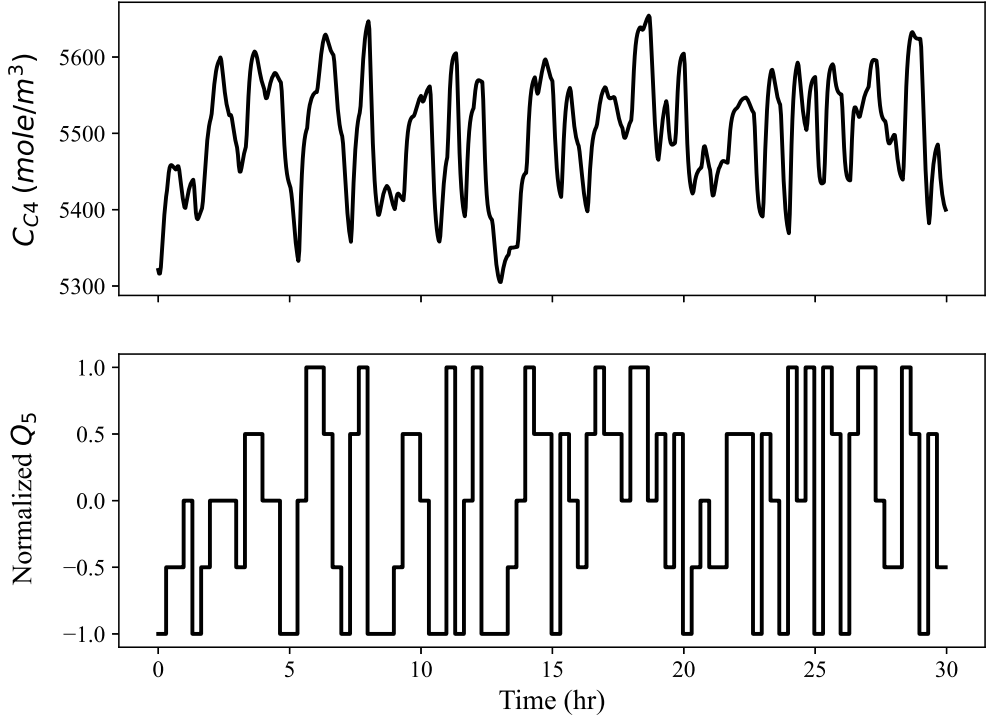}
	\caption{Multi-level input and the corresponding output response.}
	\label{mli_inout_A}
	% Thicker lines
	% \Model Reduction\Alkylation\updated model\multi_level_input_simulation.py
\end{figure}

Figure \ref{mli_inout_A} shows a portion of the input signal employed for the alkylation process and the corresponding output response. Note that the input signal is normalized with respect to the steady-state. 

At a given time step $k$, each element in the input vector $u$ randomly takes a value from a predetermined set $\mathbb{U}_d$ and hold onto that value until time $(k+n_{hold})$. Note that $n_{hold} \in \mathbb{I}_a^b = \{a, a+1, ..., b\}$, where $\mathbb{I}_a^b$ is an integer set with $a$ and $b$ being the lower and upper bounds. At time step $k+n_{hold}$, the element values of the input vector $u$ will be reselected from $\mathbb{U}_d$. 

The sets $\mathbb{U}_d$ and $\mathbb{I}_a^b$ can be designed base on the dynamics of the system such that the frequency and magnitude of the step changes can excite the system properly. If only the upper and lower bounds of the input vectors are included in $\mathbb{U}_d$, the input signal becomes a pseudorandom binary sequence (PRBS). For the alkylation process to be specific, the set $\mathbb{U}_d$ was designed to have five elements which were obtained by equally splitting the input domain $\mathbb{U}$:
\begin{equation*}
	\mathbb{U}_d = \{\underbar{u}, \underbar{u}+\frac{\delta_u}{4},  \underbar{u}+\frac{\delta_u}{2} , \underbar{u}+\frac{3\delta_u}{4}, \bar{u}\}, \: \delta_u  = \bar{u} - \underbar{u}
\end{equation*}
Taking into account of the rapid ongoing chemical reactions in the alkylation process, the integer set $\mathbb{I}_a^b$, $a = 10, b = 16$ was used. This means that the input will hold onto the same values for 20 to 32 minutes before the next set of values can be picked from $\mathbb{U}_d $.

%The dataset employed by the first-principle-model-based model reduction was generated by simulating the original system under a pseudorandom binary sequence (PRBS) signal. 

To ensure reasonable results from the considered methods, 5000 data points were generated. The reduced model order $k$ was determined by guess and check, where reduced models of different orders were generated based on the original model first and open-loop prediction performances were checked and compared through a step test. The model order that gives a satisfying balance between prediction accuracy and model order was selected and employed for the TPWL-POD models.

Data scaling and reconstructions are required for NN identification such that the preselected input-output dimension is satisfied. The Python library Keras \cite{Keras} was used to identify a fully-connected NN model which takes the form of:
\[f:\mathbb{R}^{359}\rightarrow\mathbb{R}^{15} \]
where $359 = 30 \times 8 + 14 \times 7$, 8 is the number of inputs plus outputs, 30 is the number of past data points used, 7 is the number of inputs, 14 is the number of future inputs encountered. The model takes 30 points from the past plus 17 future inputs and predicts system outputs 15 steps into the future (including the current step). Note that the number of predictions obtained by the model is the same as the length of the prediction horizon( $N = 15$), meaning that the NN needs to be called only once in the optimization problem. This helps to reduce the prediction offset caused by identifying a single-step-ahead prediction NN and calling it recursively. As for activation functions, \texttt{`sigmoid'} and \texttt{`swish'} were used in this work. Note that \texttt{`sigmoid'}  function leads to smooth nonlinear NN dynamics, while \texttt{`swish'} captures steep changes better but is more aggressive at the same time.

All simulations were done in Python. Different programming libraries were employed for different model approximation methods. To be specific, MPCTools \cite{MPCTools} was used for solving TPWL/POD-model-based EMPC, CasADi \cite{CasADi} was used for solving subspace-model-based EMPC and NN-model-based EMPC. Note that the optimization solver IPOPT \cite{IPOPT} was employed for all EMPC simulations. 

%IPOPT stands for Interior Point OPTimizer, which is a widely used numerical solver that solves large scale nonlinear optimization of continuous systems. Both MPCTools and CasADi are developed based on the IPOPT with upper-level tools added to improve the computational efficiency. To be specific, MPCTools was designed for solving nonlinear MPC problems, while CasADi requires the system to have explicit expressions, which is the reason why explicitly expressed NN is needed. 

\subsection{Results and discussion}
\subsubsection{Approximation models}
Six models are developed to approximate the nonlinear dynamics of the process. First, two TPWL models are developed by linearizing the original nonlinear model around 3 and 5 points respectively. Then, POD method is applied to the TPWL models, leading to two POD-TPWL models. One LTI model that takes the form of (\ref{subspace}) is developed using the subspace identification method. The dimension of the LTI model is 2. A fully-connected NN with two hidden layers is generated. The hidden layers have 10 and 3 nodes individually.

\begin{figure}[!t]
	\centering
	\includegraphics[width=1\columnwidth]{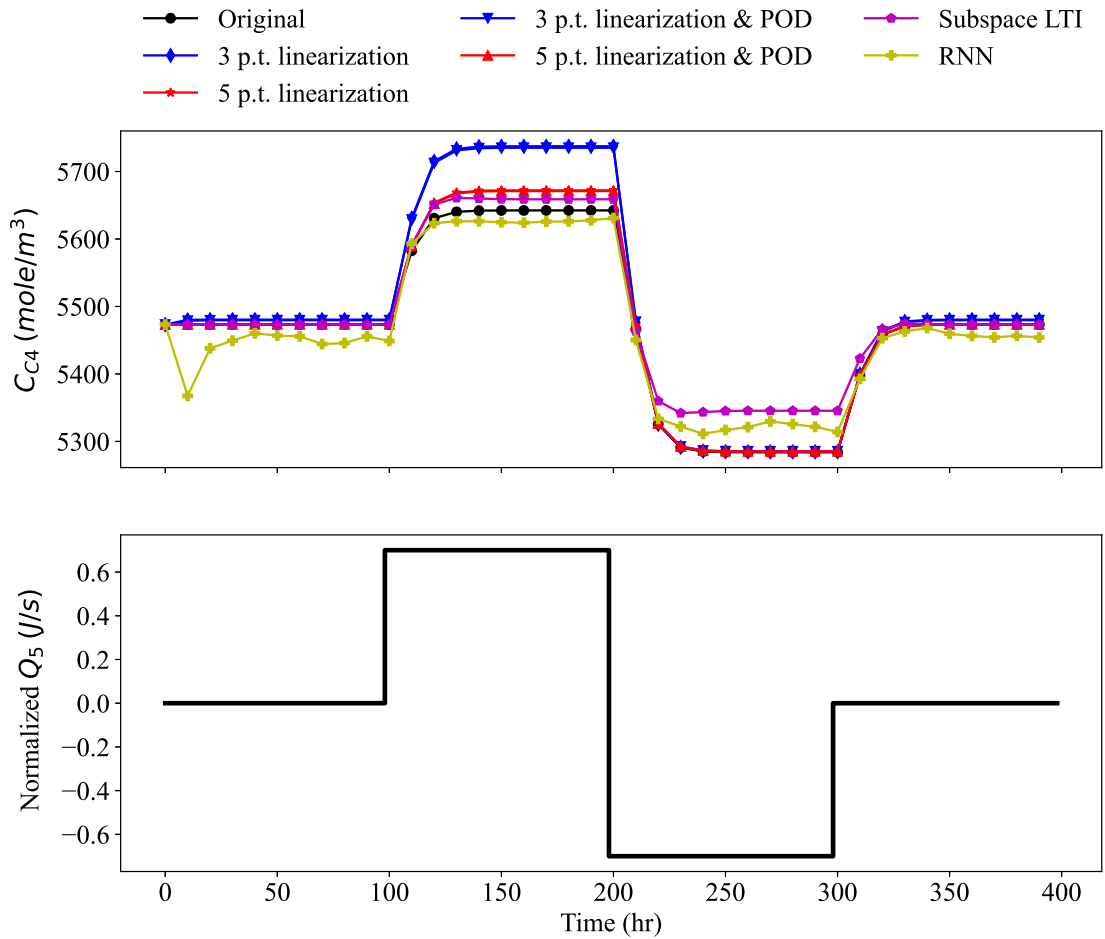}
	\caption{Open-loop output responses of the alkylation first-principle model and approximation models.}
	\label{Open_loop_A}
	% \Model Reduction\Alkylation\updated model\multi_level_input_simulation.py
\end{figure}

The original model is used as the reference for performance evaluation of the approximation models. The output responses and the corresponding input trajectory of the original model and the approximation models are shown in Figure \ref{Open_loop_A}. The prediction performance is measured by the normalized root mean square error (NRMSE):
\begin{equation}
	NRMSE = \frac{1}{y_s}\sqrt{\frac{\sum_{i=1}^{n} (\hat{y_i} - y_i)^2}{n}}
\end{equation}
where $y_s$ is the optimal steady-state output, $n$ is the number of data points, $\hat{y_i}$ is the $i$th output predicted by the approximation model and $y_i$ is the $i$th output obtained from the original nonlinear model. The NRMSE of the approximation models are summarized in Table \ref{OL_A}.

From Figure \ref{Open_loop_A}, it is observed that the TPWL and POD-TPWL models with five linearization points perform better than models with three linearization points. The application of the POD method did not impact the prediction performance significantly. Table \ref{OL_A} also shows consistent results. It was noticed that models perform differently with different variations in inputs. The NN model has a more noisy response compare to other models. The reason for this is that NNs are complex nonlinear functions with a significant number of parameters. Thus it is impractical to obtain the absolute optimal set of parameters, making numerical prediction errors hard to eliminate. Furthermore, it is noteworthy that NNs works with scaled variables in the range of -1 to 1, leading to more numerical errors. All the models respond to input changes without delays and converge to values that are very close to the steady-state output of the original nonlinear system when the input returns back to the steady-state value.  

\begin{table}[!t]
	\small
	\centering
	\caption{Open-loop prediction performance of approximation models - Alkylation}	
	%\doublerulesep 20pt
	\renewcommand\arraystretch{2}
	\label{OL_A}
	\tabcolsep 30pt
	
	\begin{tabular*}{\textwidth}{ccc}\hline
		Model applied & Dimension& Normalized RMSE\\ \hline
		3 p.t. linearized & 25 & 0.008\\
		5 p.t. linearized & 25 & 0.002\\
		3 p.t. linearized \& POD & 15 & 0.008\\
		5 p.t. linearized \& POD & 15 & 0.002\\ 
		Subspace LTI & 2 & 0.005\\ 
		Neural Network & N/A & 0.005\\  \hline
\end{tabular*}\end{table}

\subsubsection{Conventional tracking model predictive control}
\begin{figure}[!t]
	\centering
	\includegraphics[width=0.8\columnwidth]{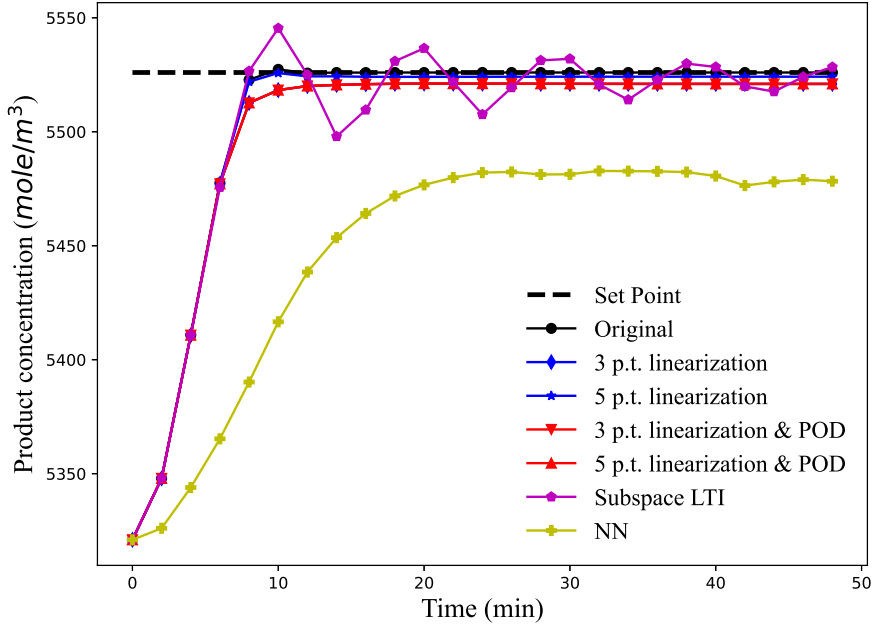}
	\caption{The optimal product concentration trajectory under different model-approximation-based conventional MPC frameworks.}
	\label{Product_C_MPC}
\end{figure}

\begin{figure}
	\centering
	\includegraphics[width=0.8\columnwidth]{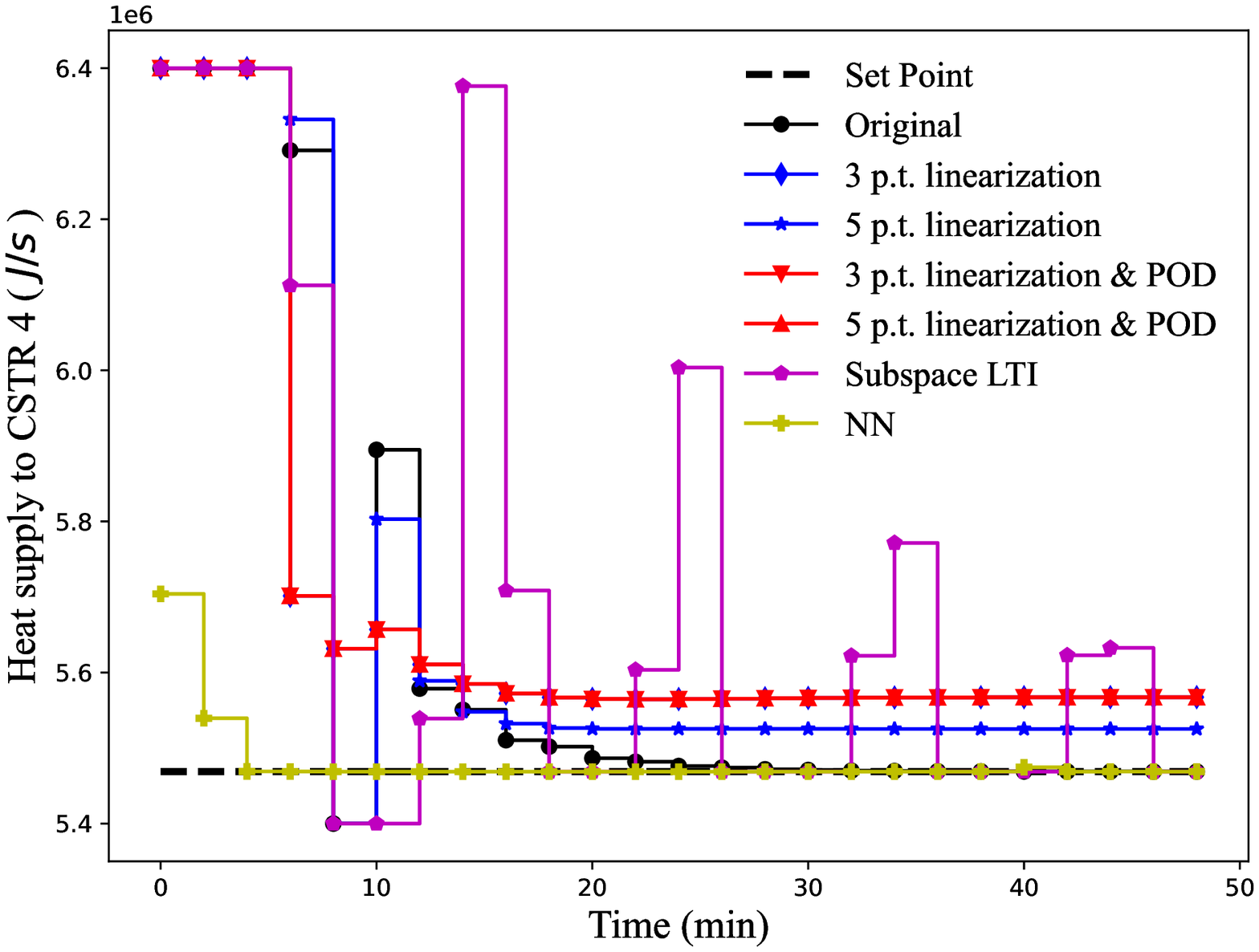}
	\caption{Manipulated input trajectory under different model-approximation-based conventional MPC frameworks - the heat supply going into CSTR-4}
	\label{Q5_MPC}
\end{figure}

The approximation models discussed in the previous section are employed by the MPC controller (\ref{MPC}) with parameter settings as discussed in Section \ref{control setting}. The simulation results are summarized in Table \ref{MPC_A}. The order of the reduced model, the average time consumed in solving the MPC problem at a given time step, and the accumulated objective function value under the corresponding trajectories are presented for the considered models. Figures~\ref{Product_C_MPC} and \ref{Q5_MPC} present the tracking performance of the system output (product concentration) and the optimal control strategy of the heat supply going into CSTR-4, which is one of the manipulated inputs. The dashed black line represents the tracking set-point, while the tracking performances obtained with different approximation models are differentiated by colors and markers. 
%Note that the performance indexes (the accumulated objective in Table~\ref{MPC_A} and the product concentration in Figure~\ref{Product_C_MPC} are all calculated using the original model for consistency. 

\begin{table}[!t]
	\small
	\centering
	\caption{MPC results of the original model and approximation models - Alkylation}	
	%\doublerulesep 20pt
	\renewcommand\arraystretch{2}
	\label{MPC_A}
	\tabcolsep 8pt
	
	\begin{tabular*}{\textwidth}{cccc}\hline
		Model applied & Dimension & Time (single step EMPC) & Objective function value \\ \hline
		Original & 25 & 1.16 & 23564.71\\
		3 p.t. linearized & 25 & 4.28 & 15619.71\\
		5 p.t. linearized  & 25 & 2.36 & 22655.89\\
		3 p.t. linearized \& POD & 15 & 0.52 & 46990.54\\
		5 p.t. linearized \& POD & 15 & 0.73 & 48512.75\\ 
		Subspace LTI & 2 & 0.07 & 49933.00\\ 
		Neural Network & N/A & 0.27 & 182511.67\\  \hline
\end{tabular*}\end{table}

As can be observed in Table \ref{MPC_A}, TPWL models failed to improve the average computational speed for solving the optimization problem however provide satisfying tracking results. This could be caused by the complexity of the TPWL models, leading to more iterations required before the numerical solver converges to the optimal solution. The increased number of linearization points does not improve MPC controller performance. A small offset in the manipulated variable can be observed in Figure \ref{Q5_MPC}. Computational speed improvements were observed with model order reduction applied on top of TPWL models while the tracking performance became worse. The observation from Figure \ref{Product_C_MPC} matches with the results in Table \ref{MPC_A}, where slight offsets in tracking were observed for the TPWL-POD models. This is reasonable since applying model order reduction and linearization at the same time further reduces the model accuracy. 

The subspace-model-based MPC provides significant computational cost reduction with an under-damping response as presented in Figures \ref{Product_C_MPC} and \ref{Q5_MPC}. It is also noticed that the control strategy obtained with the subspace model changes rapidly and more aggressively. This observation is reasonable as the subspace model is linear. Compare to nonlinear models that gives predictions that are smoother and closer to the actual dynamics of the system, the linear model naturally leads to more aggressive prediction and thus more aggressive controller responses. 

As for the NN model, a more significant offset was observed. The offset could be due to the input-output normalization required during training as well as prediction applications, leading to significant numerical errors. The non-smooth noisy prediction of the model also contributed to the offset. Computational cost reduction better than the first-principle-model-based models and worse than the subspace model was achieved. Note that the presented input happened to converge to the set-point for the NN model, which is not the case for all input variables. 

\subsubsection{Economic model predictive control}

\begin{table}[!t]
	\small
	\centering
	\caption{EMPC results of the original model and approximation models - Alkylation}	
	%\doublerulesep 20pt
	\renewcommand\arraystretch{2}
	\label{EMPC_A}
	\tabcolsep 8pt
	
	\begin{tabular*}{\textwidth}{cccc}\hline
		Model applied & Dimension & Time (single step EMPC)  & Objective function value \\ \hline
		Original & 25 & 0.68 & -6867.40\\
		3 p.t. linearized & 25 & 0.67 & -4244.25\\
		5 p.t. linearized  & 25 & 0.88 & -6840.55\\
		3 p.t. linearized \& POD & 15 & 0.60 & -3012.19\\
		5 p.t. linearized \& POD & 15 & 1.39 & -6825.65\\ 
		Subspace LTI & 2 & 0.06 & -6739.46\\ 
		Neural Network & N/A & 0.52 & -5396.95\\  \hline
\end{tabular*}\end{table}

\begin{figure}
	\centering
	\includegraphics[width=0.8\columnwidth]{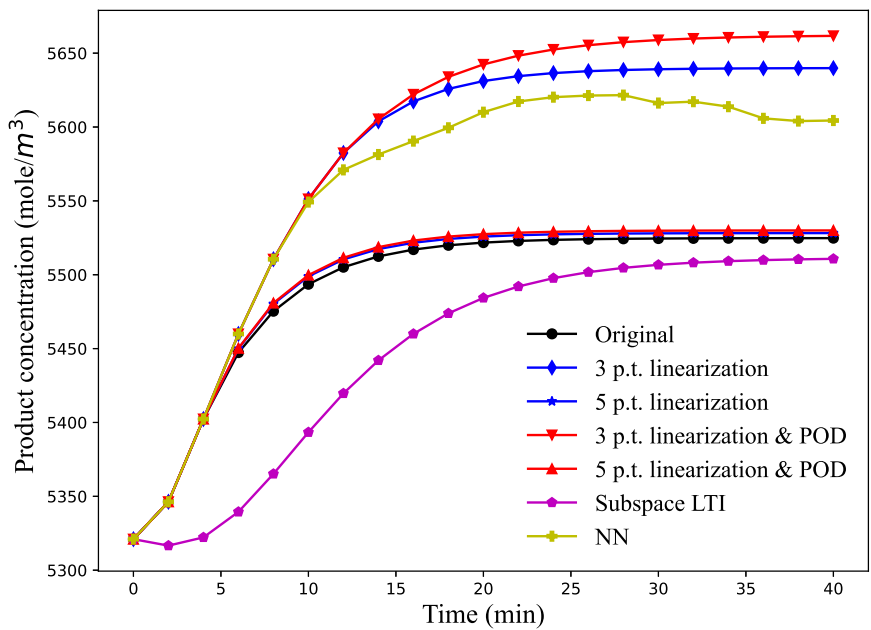}
	\caption{The optimal product concentration trajectory under different model-approximation-based EMPC frameworks.}
	\label{Product_C}
\end{figure}

\begin{figure}
	\centering
	\includegraphics[width=0.8\columnwidth]{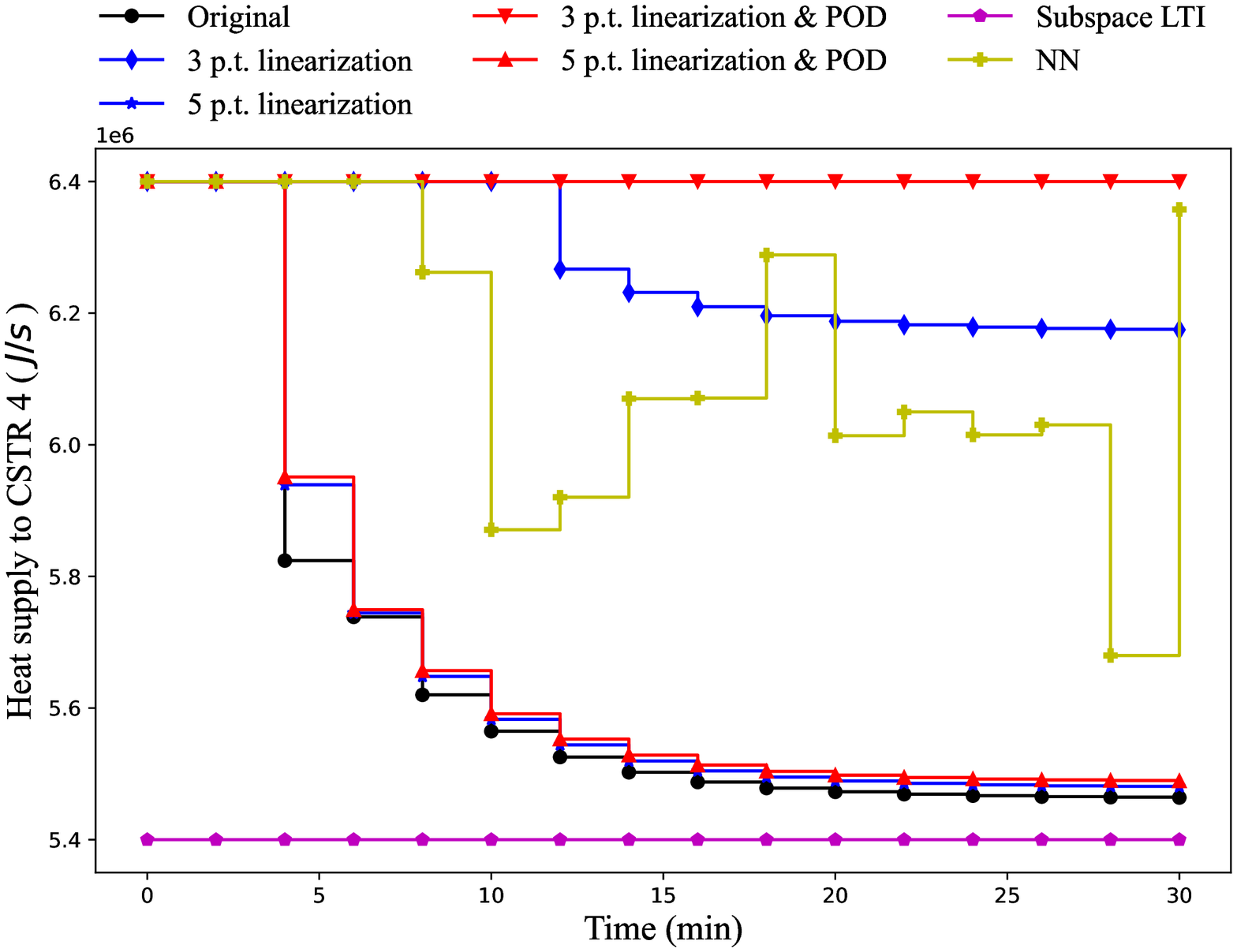}
	\caption{Manipulated input trajectory under different model-approximation-based EMPC frameworks - the heat supply going into CSTR-4}
	\label{Q5}
\end{figure}

In this subsection, we study the performance of the different models in EMPC (\ref{EMPC}). The EMPC simulation results of the alkylation process are summarized in Table~\ref{EMPC_A}. Figure~\ref{Product_C} presents the product concentration under the optimal control trajectory obtained under different model-approximation-based EMPC frameworks, of which an example is shown in Figure~\ref{Q5}, where the trajectory of the heat supply for CSTR-4 is presented. 

The TPWL model with 5 linearization points gives the best performance. Significant performance improvements are observed with increased linearization points regardless of the presence of model order reduction, which is different from the observation made for MPC controller design. It was also observed that the computational cost rises as the number of linearization points increases. Applying POD on top of TPWL leads to worse EMPC performance, which is similar to what has been observed in the MPC simulation results. When the reference model is relatively simple, further applying POD reduces the computational cost. The opposite is observed with increased reference model complexity (i.e when more linearization points are used). Furthermore, the performance deviates more significantly when the referencing LPV model is less accurate. Note that from Figure~\ref{Product_C}, it can be observed that if only the product concentration is considered, different conclusions regarding simulation performance will be obtained, as the heating cost index is ignored. However, it can still be observed from Figure \ref{Product_C} and \ref{Q5} that the 5-point-TPWL model gives control strategy that is closest to that obtained with the original model. 

Overall, no significant improvement in computational speed is achieved from using first-principle-model-based model reduction. The reason is that the complexity of the original system is relatively low, which can be proved by the fast computational speed while solving the EMPC problem with the original model. In the following section, it will be shown that the speed improvements with the usage of reduced models are significant for the more complex WWTP system.

As can be observed from Table~\ref{EMPC_A}, the computational load reduction with the subspace model employed is the highest out of all models considered. Two factors contributed to this result, namely the significant reduction in the model order and the linearity of the subspace model. A reasonable EMPC performance was also achieved with the model. 

The optimal control trajectory obtained with the NN model is similar to that obtained with the models with 3 linearization points. The response is less smooth and oscillations in the control strategy can be observed from Figure \ref{Q5}. The improvement in computational speed achieved is slightly less than the subspace LTI model but is still quite significant. The overall performance of the NN model is better than models with 3 linearization points, but worse than models with 5 linearization points and the subspace LTI model. 

%It is worth pointing out that as can be seen from Figure \ref{Q5}, the optimal control trajectory obtained with highly simplified models (e.g. the subspace LTI model) is at a steady-state over the whole simulation horizon. This is non-ideal for an EMPC framework, as the significance of the framework is that it helps to reduce economic loss during transient by tuning the manipulated inputs. A constant control strategy essentially the same as a straight set-point changing makes the significance of the advanced control framework less obvious. 

\section{Case Study 2: Wastewater treatment Plant}

\subsection{Process overview}

\begin{figure}
	\centering
	\includegraphics[width=0.8\columnwidth]{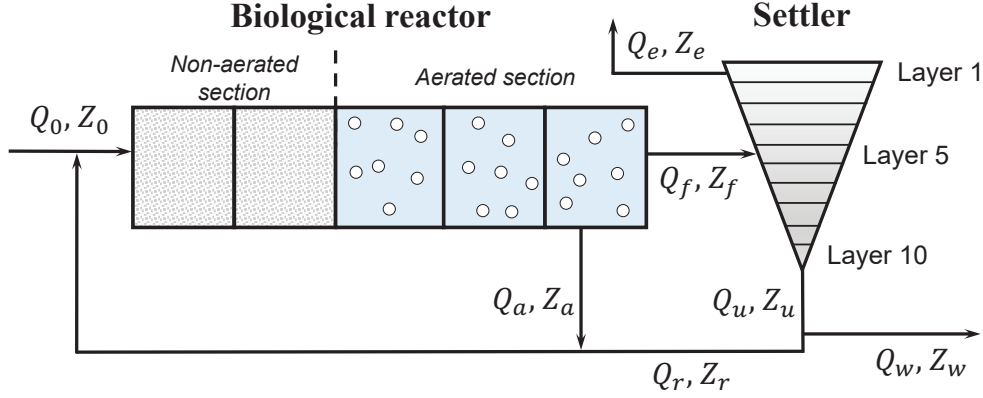}
	\caption{A schematic of the wastewater treatment plant.}
	\label{WWTP}
\end{figure}
In this section, we consider a wastewater treatment plant which is a more complex nonlinear system. A schematic of the WWTP process is shown in Figure~\ref{WWTP}. The WWTP process consists of an activated sludge bioreactor with five compartments and a secondary settler with ten layers. The reactor consists of two sections, being the non-aerated section including the first two chambers, and the aerated section which includes the latter three chambers. The wastewater stream enters the plant with a concentration $Z_0$ and a flow rate $Q_0$. After passing through the reactor, the effluent is separated into two streams, in which one of them is recycled back into the first reactor chamber at flow rate $Q_a$ and the other is fed into the settler on the fifth layer. The stream leaving from the top of the settler contains the purified water that can be disposed, while the bottom effluent is split into a recycle stream with flow rate $Q_r$ and a waste sludge stream $Q_w$. 

A detailed description of the dynamical model of the WWTP can be found in \cite{Alex2008_BSM1}. The model considers 8 biological reactions with 13 major compounds, including 7 soluble compounds and 6 insoluble compounds. The concentrations of the 13 compounds in each of the reactor chambers are the state variables. In the settler, a variable that represents all insoluble compounds is defined, leading to 8 states for each settler layers. The dynamical first-principle model of the WWTP consists of 145 states. Two manipulated input variables are considered, namely the oxygen transfer rate in the fifth chamber of the reactor $KL_{a5}$ and the flow rate of the recirculation stream $Q_a$.

%are designed to recycle wastewater and ensure the environmental impacts are minimized. In this work, the Benchmark Simulation Model no.1 (BSM1) is employed with the detailed formulation available in \cite{Alex2008_BSM1}. The process is a high-order nonlinear system (145 states) with complex ongoing biological reactions. As shown in figure \ref{WWTP},  It is assumed that bioreactions only take place in the reactor and the settler is simply responsible nonreactive for sludge settling. 
%The wastewater stream enters the plant with a concentration $Z_0$ and a flow rate $Q_0$. After passing through the reactor, the effluent is separated into two streams, in which one of them is recycled back into the first reactor chamber at flow rate $Q_a$ and the other is fed into the settler on the fifth layer. The stream leaving from the top of the settler contains the purified water that can be disposed, while the bottom effluent is split into a recycle stream with flow rate $Q_r$ and a waste sludge stream $Q_w$. 

%The two manipulated inputs are bounded with upper and lower bounds as presented in equation (\ref{input_constraints_WWTP}).
%\begin{subequations}
%	\label{input_constraints_WWTP}
%	\begin{empheq}{align}
%		0 \ \text{m}^{3}\cdot \text{d}^{-1} &Q_a\leq 2.76690\times 10^{5}\ \text{m}^{3}\cdot \text{d}^{-1} \\
%		0 \ \text{d}^{-1} \leq &K_L a_5 \leq 270 \ \text{d}^{-1}.
%	\end{empheq}
%\end{subequations} 
\subsection{Control objective and controller setting}

The economic control objective used for the EMPC controller and the steady state optimization layer of MPC are adopted from \cite{Zeng2015_IECR}. Two performance indices are considered for assessing the economic performance of the process, namely effluent quality (EQ) and overall cost index (OCI). The economic performance measure is a weighted summation of the two performance indices as follows:
\begin{equation}
	\label{WWTP_le}
	l_e = \alpha\cdot\text{EQ} + \beta\cdot\text{OCI}
\end{equation}
where EQ ($\text{kg}\ \text{pollution}\cdot\text{day}^{-1}$) is a factor that measures the processed water  quality evaluated as the daily average of a weighted summation of various effluent compound concentration, OCI covers the factors that have significant impacts on the process operating cost including sludge production ($\text{kg}\cdot\text{day}^{-1}$), aeration energy ($\text{kWh}\cdot\text{day}^{-1}$), pumping energy ($\text{kWh}\cdot\text{day}^{-1}$), and mixing energy ($\text{kWh}\cdot\text{day}^{-1}$), $\alpha$ and $\beta$ are two weighting factors for EQ and OCI, respectively. More detailed description of the cost function can be found in \cite{Zeng2015_IECR}.

The system outputs are chosen to be the states required to calculate EQ and OCI. The observability of the system is validated by PBH test. In total, 41 outputs are used. However, with only 2 inputs, it is very challenging to track all 41 outputs. Thus the MPC controller is designed to track two of the outputs that deviate the most from the set-point, namely the slowly biodegradable and soluble substrate in the first and second reactor compartments. The two tracked outputs will be denoted as $XS_1$ and $XS_2$ in the following sections. The weighting parameters are defined to be $Q = diag([100,100])$, $R = diag([100, 100])$ and $P_f = diag([1000,1000])$. Similar to the alkylation process, the input and output vectors are only bounded:
\begin{subequations}
	\begin{empheq}{align}
		\mathbb{U} &= \{u|\: \underbar{u} \leq u \leq \bar{u}\} \\
		\mathbb{Y} &= \{y|\: y\geq 0\}
		%, \: u_{min} = [-1.0, -1.0, -1.0, -1.0, -1.0, 0.0, 0.0],  \:[1.0, 1.0, 1.0, 1.0, 1.0, 1.0, 1.0]
	\end{empheq}
\end{subequations}
where $\underbar{u} = [0, 0], \bar{u} = [240, 92230]$. Scaling is performed in implementations to reduce numerical errors. 

\begin{figure}[!t]
	\centering
	\includegraphics[width=0.8\columnwidth]{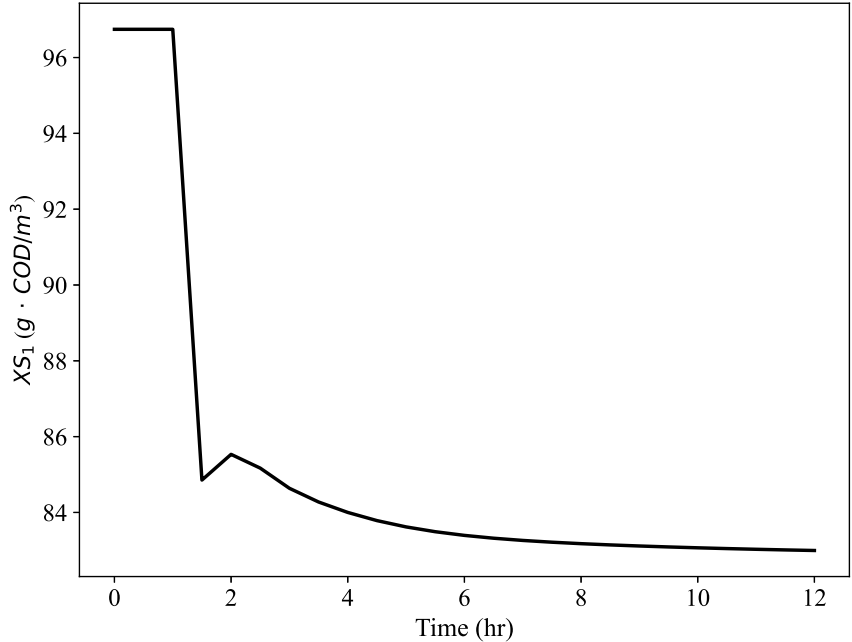}
	\caption{Step response of a WWTP process output.}
	\label{Step_WWTP}
	% C:\Users\zhiyinan\OneDrive - ualberta.ca\Model Reduction\WWTP\simulation_WWTP.py
\end{figure}

An open-loop response of $XS_1$ with respect to step changes in the input vector at $t = 1 \, hr $ is presented in Figure \ref{Step_WWTP}. The controller sampling time is 30 minutes and the prediction horizon is $N = 10$. The dynamics five hours into the future are considered by the controller, which is enough to cover the essential dynamics of the process according to Figure \ref{Step_WWTP}. 

\subsection{Data generation and model approximation}

\begin{figure}[!t]
	\centering
	%\epstopdfsetup{outdir=./}
	\includegraphics[width=0.8\columnwidth]{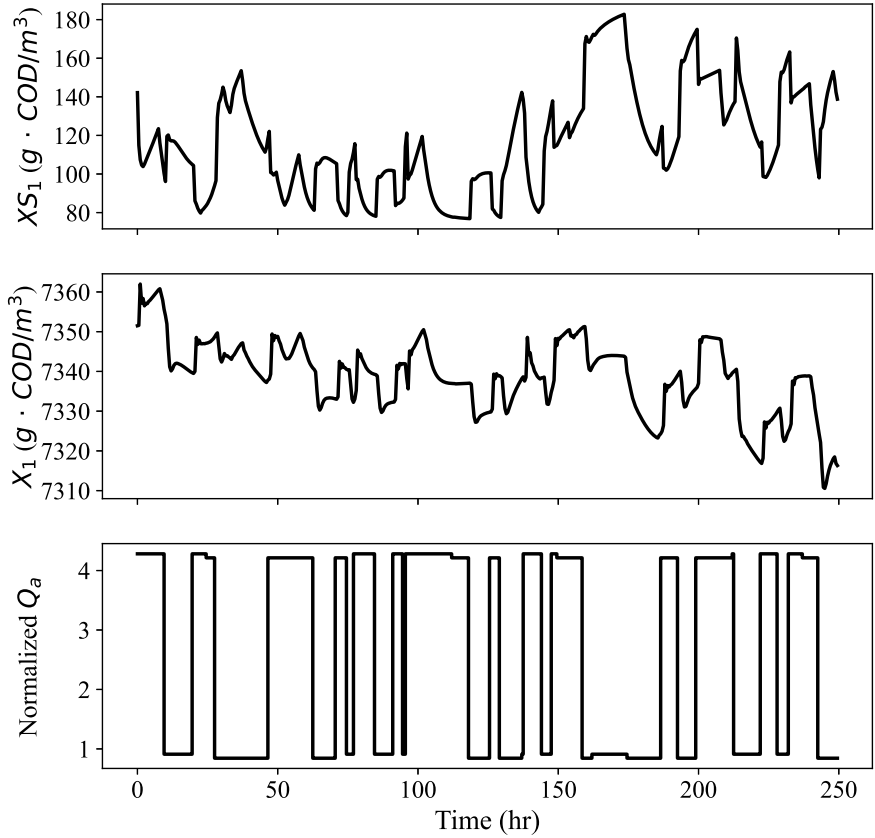}
	\caption{Multi-level input and the corresponding output response.}
	\label{mli_inout_WWTP}
	% Thicker lines
	% \Model Reduction\Alkylation\updated model\multi_level_input_simulation.py
\end{figure}

A dataset with 5376 data points (112 simulation days) is generated for the WWTP process. Samples of the training dataset are presented in Figure \ref{mli_inout_WWTP}. Two output trajectories and one input trajectory are presented. Recall that $XS_1$ is the slowly biodegradable and soluble substrate in the first reactor compartment, $Q_a$ is the flow rate of the recirculation stream. $X_1$ represents the total sludge concentration in the bottom layer of the settler. Note that the input signal is designed differently compare to the alkylation process. Instead of equally splitting the input domain, the signal is generated by adding small additional steps to a PRBS signal. This is because if the same approach is used, the complex nonlinear dynamics of the WWTP lead to an ill-conditioned dataset that can not be used for model identification. By making small changes to a working PRBS input signal, the corresponding output responses would be similar to those obtained with the PRBS signal but are able to capture the system nonlinearity. Here the small steps are added by another PRBS signal with magnitudes $[0, 0.1\times (\bar{u}_{data} - \underbar{u}_{data})]$, where $\bar{u}_{data}$ and $\underbar{u}_{data}$ are the upper and lower bounds of the PRBS signal. 

System orders of the approximation models are again determined by guess and check such that fair prediction performances are achieved. A fully-connected NN model in the following form is generated:
\[f:\mathbb{R}^{878}\rightarrow\mathbb{R}^{410} \]
where $878 = 20 \times 43 + 9 \times 2$, $410 = 10 \times 41$, 43 is the number of inputs plus outputs (2 inputs and 41 outputs), 20 is the number of past data points, 9 is the number of future inputs, 10 is the number of output steps been predicted. Again the NN is designed to fit the controller design such that the model needs to be called only once while defining the optimization problem.

%The identified RNN model is applied to the EMPC optimization problem following the same framework as the subspace model presented in the previous section. Different from the two approaches discussed before, the IPOPT solver \cite{IPOPT} is directly used for solving optimization problems due to the lack of mathematically expressed system models. The IPOPT stands for Interior Point OPTimizer, which is a widely used software library that solves the large scale nonlinear optimization of continuous systems. Both MPCTools and CasADi mentioned in the previous sections are developed based on the IPOPT with upper-level tools added to improve the computational efficiency for optimization problems with particular formulations. To solve a given optimization problem, the IPOPT requires users to provide the gradient of the objective function and the Jacobian and Hessian matrices of the constraints (if any) with respect to the decision variables. In this work, the finite difference method is used to estimate these factors. 

\subsection{Results and discussion}

\subsubsection{Approximation models}

Six approximation models are identified for the WWTP process. Based on the original model, three sets of linearization points are investigated, namely 1, 4, and 5 linearization points. Note that due to the complexity of multi-point linearized TPWL models, simulations are not performed with TPWL models with 145 states. Instead, only TPWL-POD models with 47 artificial reduced states are employed in MPC and EMPC simulations. An LTI model with state vector dimension $s = 48$ is identified using the subspace method. The fully-connected NN with input-output design mentioned in the previous section consists of two hidden layers, of which each layer contains 40 nodes. 

\begin{figure}[!t]
	\centering
	\includegraphics[width=0.8\columnwidth]{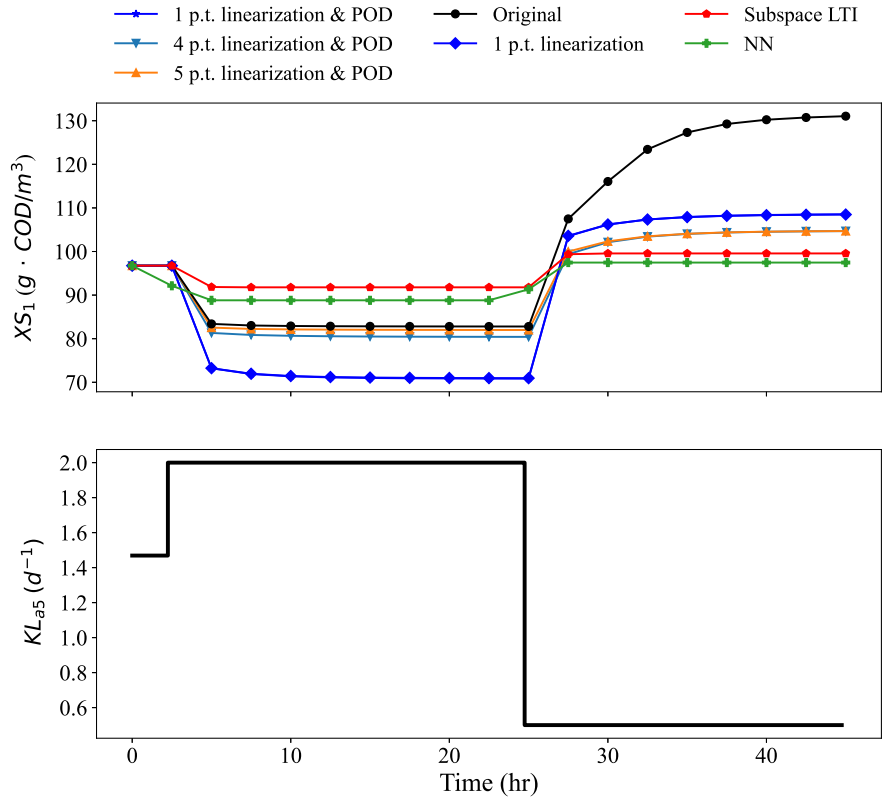}
	\caption{Open-loop output responses of the WWTP first-principle model and approximation models.}
	\label{Open_loop_WWTP}
	% \Model Reduction\Alkylation\updated model\multi_level_input_simulation.py
\end{figure}

Figure \ref{Open_loop_WWTP} presents the open-loop response of $XS_1$ and the corresponding input signal of $KL_{a5}$. The average RMSE of the outputs predicted by approximation models are summarized in Table \ref{OL_W}. The application of the POD method has little effect on the open-loop prediction performance of the model with a single linearization point. As for the TPWL-POD models, the model performance improves slightly with increased linearization points. The subspace model has the worst performance due to significant information loss caused by the model linearity. The NN model performs slightly better than the subspace model as it captures the nonlinearity of the system to some extent. From Table \ref{OL_W}, it can be concluded that the single-point-linearized model has the best prediction performance. To summarize, all approximation models respond to input changes without delay. 
\begin{table}[!t]
	\small
	\centering
	\caption{Open-loop response of the reduced model and the original model - WWTP}
	\label{OL_W}
	%\doublerulesep 20pt
	\renewcommand\arraystretch{2}
	\tabcolsep 50pt
	\begin{tabular*}{\columnwidth}{ccc}\hline
		Model applied & Order & RMSE \\ \hline
		1 p.t. linearized & 145 & 0.056\\
		1 p.t. linearized \& POD & 47 & 0.057\\
		4 p.t. linearized \& POD & 47 & 0.059\\
		5 p.t. linearized \& POD & 47 & 0.057\\ 
		Subspace LTI & 48 & 0.070\\
		%Subspace LPV & 48 & 6.94 & 127355.27\\ 
		Neural Network & N/A & 0.067\\ \hline
	\end{tabular*}
\end{table}

\subsubsection{Conventional tracking model predictive control}
\begin{figure}
	\centering
	\includegraphics[width=0.8\columnwidth]{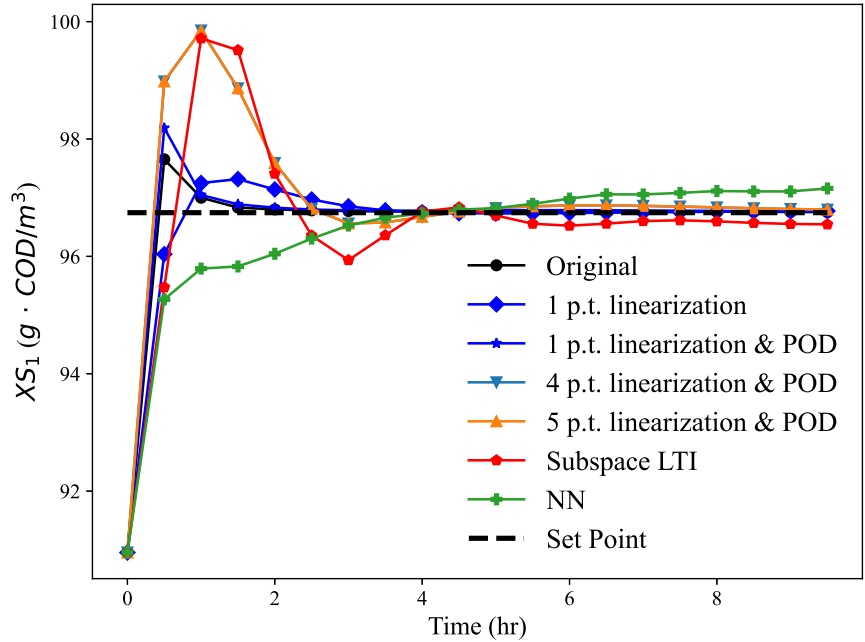}
	\caption{Optimal output trajectory of the WWTP under different model-approximation-based conventional MPC frameworks.}
	\label{XS1_MPC}
	% \Model Reduction\Alkylation\updated model\multi_level_input_simulation.py
\end{figure}

\begin{figure}
	\centering
	\includegraphics[width=0.8\columnwidth]{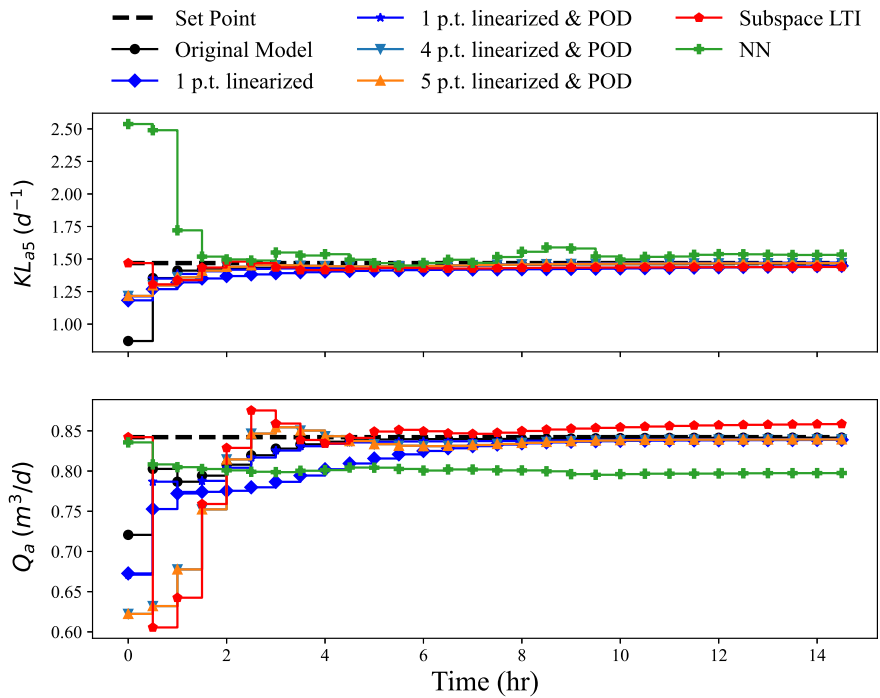}
	\caption{Optimal control policy of the WWTP under different model-approximation-based conventional MPC frameworks.}
	\label{input_MPC}
	% \Model Reduction\Alkylation\updated model\multi_level_input_simulation.py
\end{figure}

MPC simulations based on the approximation models introduced in the previous section are carried out and the results are summarized in this section. Table \ref{MPC_Results_W} presents the average computational costs and the accumulated objective function values of the original model and the approximation models. The dynamic trajectories of $XS_1$ under each of the MPC controllers are shown in Figure \ref{XS1_MPC}, while Figure \ref{input_MPC} presents the optimal control strategies.

As can be observed from Table \ref{MPC_Results_W}, the single-point linearized model with reduced order converges to the set-point fastest. According to Figures \ref{XS1_MPC} and \ref{input_MPC}, it also gives the optimal control strategy closest to that obtained with the original nonlinear model. Note that the model leads to the largest computational cost among first-principle-model-based approximation models, indicating the effect of the number of states considered in the optimization problem. The second fastest convergence to the set-point is given by the single-point linearized model with order reduction. A milder response in the output is also observed from Figure \ref{XS1_MPC}. The multi-point-linearized TPWL-POD models give similar responses, with more significant overshoots in the output at the beginning. It is also noticed that for input $Q_a$, the control strategies obtained based on these two models start further from the set-point. It is also observed that with an increased number of linearization points, the computational complexity increases.

Being the most efficient in terms of computational cost, the subspace model has the worst performance. A huge overshoot is observed at the beginning and it failed to converge to the set-point without offset. As shown in Figure \ref{XS1_MPC}, the subspace model leads to an underdamped response after the first overshooting peak. These indicate a more significant plant-model-mismatch between the subspace model and the real system. The NN model is the most computationally expensive out of all approximation models considered. This could be due to the less smooth multi-step-ahead prediction trajectory, leading to higher costs in computing Jacobian and Hessian while solving the optimization problem. It shows better performance compared to the subspace model, however offset is as well observed. Figure \ref{input_MPC} indicates the NN model leads to an significant offset for input $Q_a$. To be noteworthy, the NN model is the only model that leads to an output response without overshoot in the beginning. 
\begin{table}[!t]
	\small
	\centering
	\caption{MPC results of the reduced model and the original model - WWTP}
	\label{MPC_Results_W}
	%\doublerulesep 20pt
	\renewcommand\arraystretch{2}
	\tabcolsep 35pt
	\begin{tabular*}{\columnwidth}{cccc}\hline
		Model applied & Order & Time & Objective \\ \hline
		Original & 145 & 39.63 & 1170.32\\
		1 p.t. linearized & 145 & 4.75 & 3891.11\\
		1 p.t. linearized \& POD & 47 & 2.46 & 1403.76\\
		4 p.t. linearized \& POD & 47 & 2.78 & 4736.07\\
		5 p.t. linearized \& POD & 47 & 2.84 & 4736.07\\ 
		Subspace LTI & 48 & 0.48 & 11696.64\\
		%Subspace LPV & 48 & 6.94 & 127355.27\\ 
		Neural Network & N/A & 9.62 & 6647.17\\ \hline
	\end{tabular*}
\end{table}

\subsubsection{Economic model predictive control}

%\textcolor{red}{First, add a short paragraph to briefly discribe the different cases considered or will be compared. Indeed, you have the information in Table 1. But it is better to introduce before presenting the results.}

%\textcolor{red}{Then, say Table 1 shows the results. Also, present the two figures added. } 

%\textcolor{red}{Subsequently, discuss the results one by one. Start from linearization \& POD, then subspace, then RNN. Focus on computational complexity and economic performance. You can use some of the paragraphs you have below. Re-organize and add or improve. Make sure you read to fix the language. }

\begin{figure}
	\centering
	\includegraphics[width=0.8\columnwidth]{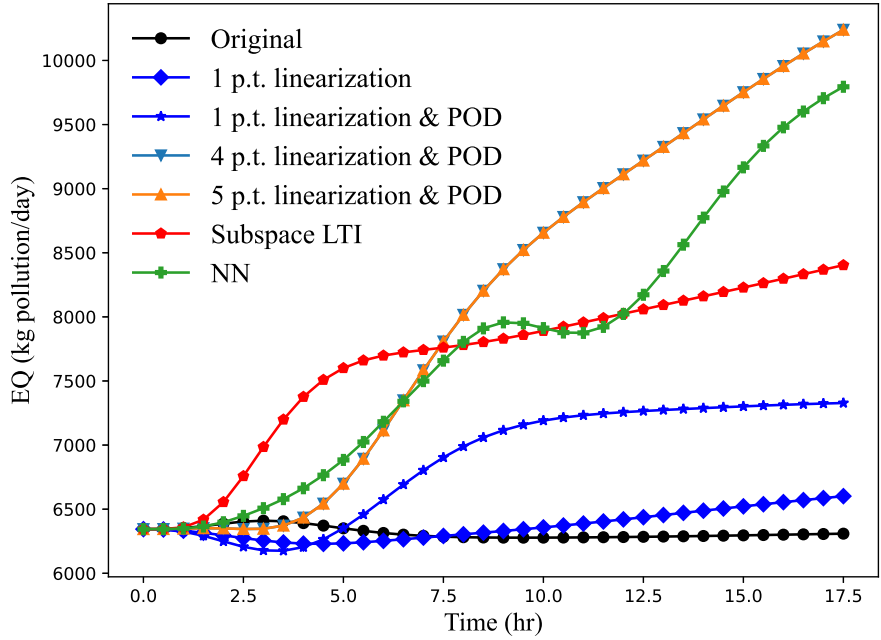}
	\caption{Effluent quality (EQ) index trajectory under different model-approximation-based EMPC frameworks. }
	%\textcolor{red}{Let Latex choose where to put the table and figures. In general, we prefer the tables and figures are at the top of a page. Also, in the two figures, the legends are not clear. Try to improve. }
	\label{EQ}
\end{figure}

\begin{figure}
	\centering
	\includegraphics[width=0.8\columnwidth]{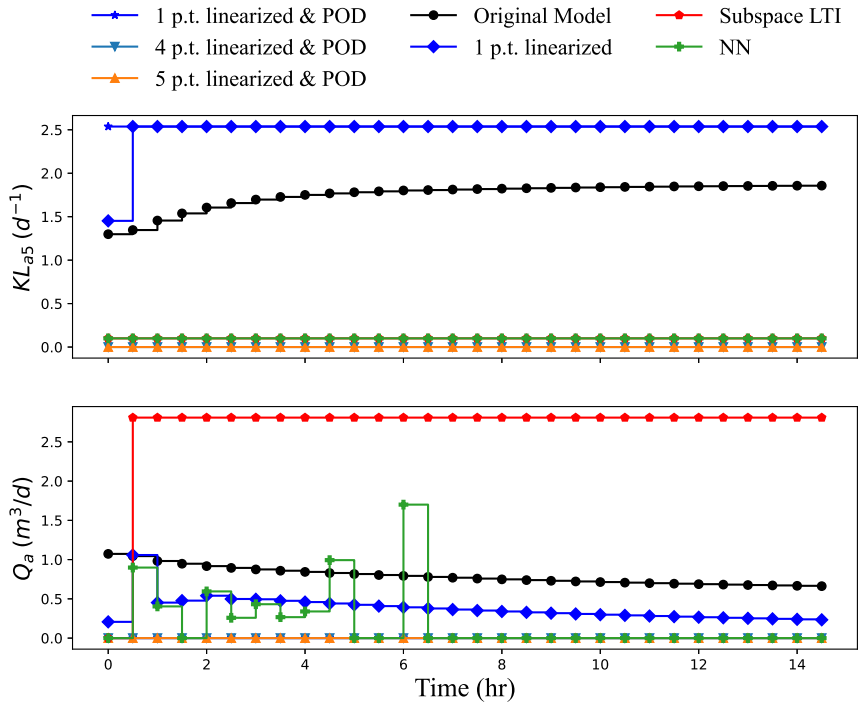}
	\caption{Optimal control trajectories under different model-approximation-based EMPC frameworks}
	\label{input}
\end{figure}

EMPC simulations are as well performed with the approximation models and the original nonlinear model.  Table~\ref{Results_W} presents the EMPC simulation results of the WWTP process. Figure~\ref{EQ} shows the trajectory of the effluent quality (EQ) index under the optimal control strategies obtained with different methods, which is presented in Figure~\ref{input}. 

%From Table~\ref{Results_W}, it can be observed that all the models considered except the RNN model improve the computational speed significantly. 
For the first-principle-model-based model reduction, it is observed from Table~\ref{Results_W} that the single-point linearized model has the best economic performance, which is the same as that observed in MPC simulations. With model order reduction, the computational speed was further improved however with a worse performance at the same time. The time required to solve the optimization problem increases steeply when the number of linearization points increases. The effect of model complexity is noticed to have a greater impact on EMPC compare to MPC. The economic performance however, did not improve when the number of linearization points increases. This could be because the system operates closely around the steady-state. It can be seen from Figure \ref{EQ} that apart from that obtained with the original model, the best effluent quality is achieved with the single-point TPWL model. The EQ index trajectories of the multi-point TPWL-POD methods are worse and similar to each other, which matches with the results as shown in Table \ref{Results_W}.  

\begin{table}[!htb]
	\small
	\centering
	\caption{EMPC results of the reduced model and the original model - WWTP}
	\label{Results_W}
	%\doublerulesep 20pt
	\renewcommand\arraystretch{2}
	\tabcolsep 35pt
	\begin{tabular*}{\columnwidth}{cccc}\hline
		Model applied & Order & Time & Objective \\ \hline
	    Original & 145 & 396.07 & 115862.71\\
		1 p.t. linearized & 145 & 7.06 & 118614.70\\
		1 p.t. linearized \& POD & 47 & 5.64 & 122521.16\\
		4 p.t. linearized \& POD & 47 & 23.68 & 138520.05\\
		5 p.t. linearized \& POD & 47 & 42.16 & 138520.05\\ 
		Subspace LTI & 48 & 6.86 & 121468.84\\
		%Subspace LPV & 48 & 6.94 & 127355.27\\ 
		Neural Network & N/A & 12.65 & 135274.59\\ \hline
	\end{tabular*}
\end{table}

The subspace model leads to EMPC performances better than the multi-point linearized models but worse than the single-point linearized model. The reduction in computational cost is more significant than that achieved with the single-point linearized model as the order of the subspace models is much lower. 

The NN-based EMPC framework provides economic performance slightly better than that obtained by the multi-point TPWL-POD models. The computational cost is lower than multi-point TPWL-POD models but higher than the LTI models. This is reasonable as NN is nonlinear by its nature. It is noticed from Figure~\ref{input} that the optimal control trajectory of $Q_a$ obtained based on the NN osculates the most out of all models considered. This may be caused by the non-smooth dynamics of the NN, leading to challenges while solving the optimization problem. 

%It is to be noted that, as observed in the first case study, the optimal control strategies of some simplified models are again constant over the horizon, which reduces the significance of using the advanced control framework.
% As the RNN model was repeatedly called during iteration, the difference in prediction cost was further increased. 

\section{Concluding remarks}

The focus of this paper is on evaluating the applicability and performance of representative model approximation methods in MPC and EMPC. While these model approximation methods have been successful in open-loop dynamic prediction, the nature of advanced control indeed poses great challenges for these existing model approximation methods as demonstrated in the simulations of this work. We have the following conclusions and recommendations:
\begin{itemize}
	\item Approximation models do not always lead to faster computation. The time required to do one iteration while solving the optimization problem might be reduced from using an approximation model. However, depending on the model structure, the number of iterations required might increase, resulting in a longer computational time.
	
	\item The same approximation method could perform very differently on different systems. For example, increased linearization points improve model performance for the alkylation process but not the WWTP process.
	
	\item For MPC, model accuracy and structure are essential for achieving satisfying set-point-tracking without offsets. More investigations are necessary for dataset-based approximation approaches. 
	
	\item EMPC tries to track the trajectory that maximizes the economic performance. This requires the model has sufficient accuracy over the entire operating range, which poses challenges for the existing methods. 
	
	\item The computational complexity of POD combined with linearization increases quickly with the increase of the number of linearization points. At the same time, the increased number of linearization points does not necessarily lead to improved control performance in the examples. How to determine the optimal number of linearization points may be a question worth further investigation. 
	%\item In general, using the POD method alone cannot improve the computational speed, because calculations from the original nonlinear model are still required.
	%\item For systems that operate closely around the steady-state, the single point linearization outperforms the multi-point linearization.
	%\item While LTI subspace identification has been used in many applications, the LPV version subspace identification seems not that mature, yet. This may make its application in EMPC of nonlinear systems challenging. 
	
	\item  The NN models identified in this work focus on multi-step-ahead prediction. These models may give sufficiently accurate approximations of nonlinear systems.  However, oftentimes these models return non-smooth prediction trajectories, which may pose challenges in the associated optimization. More work needs to be done to improve the overall framework. For example, an optimal operating point can be determined with NN-based EMPC first, followed by designing a more accurate controller based on the obtained set-point for better control strategies.
	
	%\item As summarized above, while NN potentially may approximate a nonlinear system with sufficient accuracy, it may pose challenges in the associated optimization. 
	
\end{itemize}

%Based on the known original model, it is found that the performance of model reduction methods strongly depends on the nature of the original system. For highly-nonlinear systems, the performance with LPV reduced models improves with the number of linearization points used. The opposite can be observed for systems with low-nonlinearity, of which LTI reduced model gives the best performance. The POD method only reduces the computational complexity when the system order reduced is significant and performs better if the interested model has high accuracy.  

\end{document}